\documentclass[a4paper,fleqn,usenatbib]{mnras}

\usepackage{mathptmx}

\usepackage[T1]{fontenc}
\usepackage{ae,aecompl}

\usepackage{graphicx}	
\usepackage{amsmath}	
\usepackage{amssymb}	
\usepackage{pdflscape}

\title[Classifier to Detect Elusive Astronomical Objects]{A Classifier to Detect Elusive Astronomical Objects through Photometry}

\author[]{
D. Bhavana$^{1}$,
S. Vig$^{1}$\thanks{Email: sarita@iist.ac.in},
S.K. Ghosh$^{2}$,
and Rama Krishna Sai S. Gorthi$^{3}$\thanks{E-mail: rkg@iittp.ac.in}
\\
$^{1}$Indian Institute of Space science and Technology, Thiruvananthapuram 695547, India\\
$^{2}$Tata Institute of Fundamental Research, Mumbai 400 005, India\\
$^{3}$Indian Institute of Technology, Tirupati 517506, India
}

\date{Accepted XXX. Received YYY; in original form ZZZ}

\pubyear{2018}

\begin{document}
\label{firstpage}
\pagerange{\pageref{firstpage}--\pageref{lastpage}}
\maketitle

\begin{abstract}
The application of machine learning principles in the photometric search of elusive astronomical 
objects has been a less-explored frontier of research. Here we have used three methods: the Neural Network and two variants of k-Nearest Neighbour, to identify brown dwarf candidates using the photometric colours of known brown dwarfs. We initially check the efficiencies of these three classification techniques, both individually and collectively, on known objects. This is followed by their application to three regions in the sky, namely Hercules ($2^\circ\times2^\circ$), Serpens ($9^\circ\times4^\circ$) and Lyra ($2^\circ\times2^\circ$). Testing these algorithms on sets of objects that include known brown dwarfs show a high level of completeness. This includes the Hercules and Serpens regions where brown dwarfs have been detected. We use these methods to search and identify brown dwarf candidates towards the Lyra region. We infer that the collective method of classification, also known as ensemble classifier, is highly efficient in the identification of brown dwarf candidates.
\end{abstract}

\begin{keywords}
methods: statistical -- (stars): brown dwarfs --  infrared: stars -- techniques: miscellaneous -- techniques: photometry
\end{keywords}



\section{Introduction}

Classification of astronomical objects has always posed a problem to researchers. Classification usually depends on characteristic spectral features of a set of objects that can be observed through photometry at certain wavelengths in the absence of spectroscopic data. Such photometry-based classification schemes have been traditionally implemented by applying colour and magnitude cuts to the data \citep{Veen1988,Allen2004}. The advent of large all-sky photometric surveys, each identifying millions of new objects, necessitates automated techniques for classification. It is clear that increasing the number of features used for classification improves accuracy. The need for handling multi-dimensional feature spaces (wherein different classes can be clearly distinguished) has led to the introduction of machine learning algorithms for this purpose \citep{Agrawala1968}. 

Among the earliest machine-learning methods used for classification in astronomy are the \texttt{k}-Nearest Neighbour (\texttt{k}-NN) method and the Neural Network (NeuN) algorithm. These computational techniques for recognising patterns have been used for astronomical classification since 1970s. A statistical nearest-neighbour test was employed by \citet{Bogart1973} to study clustering of galaxies and QSOs. \citet{Heydon-Dumbleton1989} used an automatic classification procedure for star-galaxy classification using NeuN. Thereafter, \citet{Odewahn1992} used the perceptron and backpropagation neural network algorithms to create accurate classifiers for separating star and galaxy images. Similar networks were also used for morphological classsification of galaxies by \citet{StorrieLombardi1992}. Nowadays, more sophisticated techniques are used for classification, such as the Support Vector Machines (SVMs) \citep{Krakowski2016,Kurcz2016}, Random Forest algorithm \citep{Nakoneczny2018} and deep learning \citep{Gonzalez2018}. In the current work, we attempt to create an ensemble classifier by applying simple machine learning techniques like k-NN and NeuN for identification of objects like brown dwarfs, which are rarely observed despite being theoretically predicted to be in abundance \citep{Muvzic2017}.

Brown dwarfs are objects with mass so low that they cannot sustain hydrogen fusion in their cores. They are capable of fusing deuterium, and the minimum mass for deuterium-burning is defined as the lower mass limit for a brown dwarf \citep[adopted by the International Astronomical Union in 2002,][]{Spiegel2011}. Brown dwarfs fall under 3 spectral types, L, T and Y. Some of the hottest brown dwarfs discovered have been found to be of late M-type as well, but since the photometric properties of M-type brown dwarfs are very similar to the M-type main-sequence stars, we have omitted them from the brown dwarf category in this study.
Brown dwarfs are extremely faint and cool; T and Y dwarf temperatures can range from 300 K to 1300 K, and visible magnitudes of even the hotter L-dwarfs fall above 20 mag \citep{Costa2005}. Hence, they are very difficult to detect. Their emission peaks in the infrared with a distinctive spectral energy distribution arising from strong molecular absorption features \citep{warren2007,Stephens2009}. Foraging the immediate solar neighbourhood for such cold objects is one of the goals of the all-sky mission performed by the Wide-field Infrared Survey Explorer \citep[WISE;][]{Wright2010}. WISE is an Earth-orbiting NASA mission that surveyed the entire sky simultaneously at wavelengths of 3.4, 4.6, 12, and 22 \micron , hereafter referred to as bands W1, W2, W3, and W4, respectively. These bands have been selected so as to uniquely identify brown dwarfs based on their spectral features \citep{Kirkpatrick2011}. 

Most of the brown dwarf searches till date have used generic colour cuts to identify candidates, 
which are then confirmed or rejected on the basis of additional spectroscopy and proper motion studies \citep{Cushing2011,Kirkpatrick2011,Tinney2012,Mace2013}. However, \citet{Luhman2012} cautions against the use of theoretical magnitudes and colours for this purpose. Instead, he recommends the use of photometry of known low-mass objects, to guide the identification of candidates. But this has been attempted by very few as of now. For example, \citet{Marengo2009} present a statistical method for the photometric search of rare astronomical sources using the \texttt{k}-NN method. Here \texttt{k}-NN acts as a non-parametric classifier, by deciding the class of a new set of data based on its distance from a class of brown dwarf templates, where the distance is defined in multi-dimensional colour and magnitude space. We have included the above technique, hereby termed as \texttt{k}-NN threshold distance, in our analysis. In addition, we have applied a modified version of \texttt{k}-NN distance method, namely \texttt{k}-NN classification, whereby additional classes (i.e. different sets of background objects which are not brown dwarfs) are considered for improvement in the analysis. The third method used in the present work is NeuN, a parametric classifier. Although this method has been used for astronomical classification problems as mentioned earlier, it has not been applied to the specific problem of brown dwarf classification problem till now. 

In this work, we specifically aim to (i) pose the identification of brown dwarfs as a classification problem rather than a (colour) threshold based identification, (ii) identify satisfactory training data for brown dwarfs as well as background classes, and (iii) examine a suitable approach for identifying brown dwarf candidates using their infrared photometry by applying well-known methods like \texttt{k}-NN classifier and NeuN. In addition, we propose an ensemble classifier for the same. The ensemble classifier will identify the final brown dwarf candidates on the basis of a majority vote, with each individual classifier being given equal weightage. Simultaneously, we also use different training sets and compare their efficiencies to determine an optimum set.

The organisation of this paper is as follows. Section~\ref{sec:Sec2}  of this paper describes the machine learning methods, while Sect.~\ref{sec:Sec3} provides details of the data used and features employed to form the training sets.  In Sect.~\ref{sec:Sec4}, the outcome of the classification techniques are determined through efficiencies of few optimum training sets. These training sets are applied to certain regions in sky in Sect.~\ref{sec:Sec5} and inferences drawn in Sect.~\ref{sec:Sec6}. A short summary of results is provided in Sect.~\ref{sec:Sec7}.   

\section{Classification Schemes}
\label{sec:Sec2}
The data used for the photometric classification include the colours of objects as derived from near and mid-infrared photometric bands. The colours are designated as features in machine-learning parlance, and these are described in Sect.~\ref{sec:colours}. Objects are classified based on their photometric similarity to known objects of a given class. These latter objects are referred to as templates. Below, we briefly introduce the classification schemes.

\subsection{Neural Networks}
\begin{figure}
	\centering
	\includegraphics[width=0.30\textwidth]{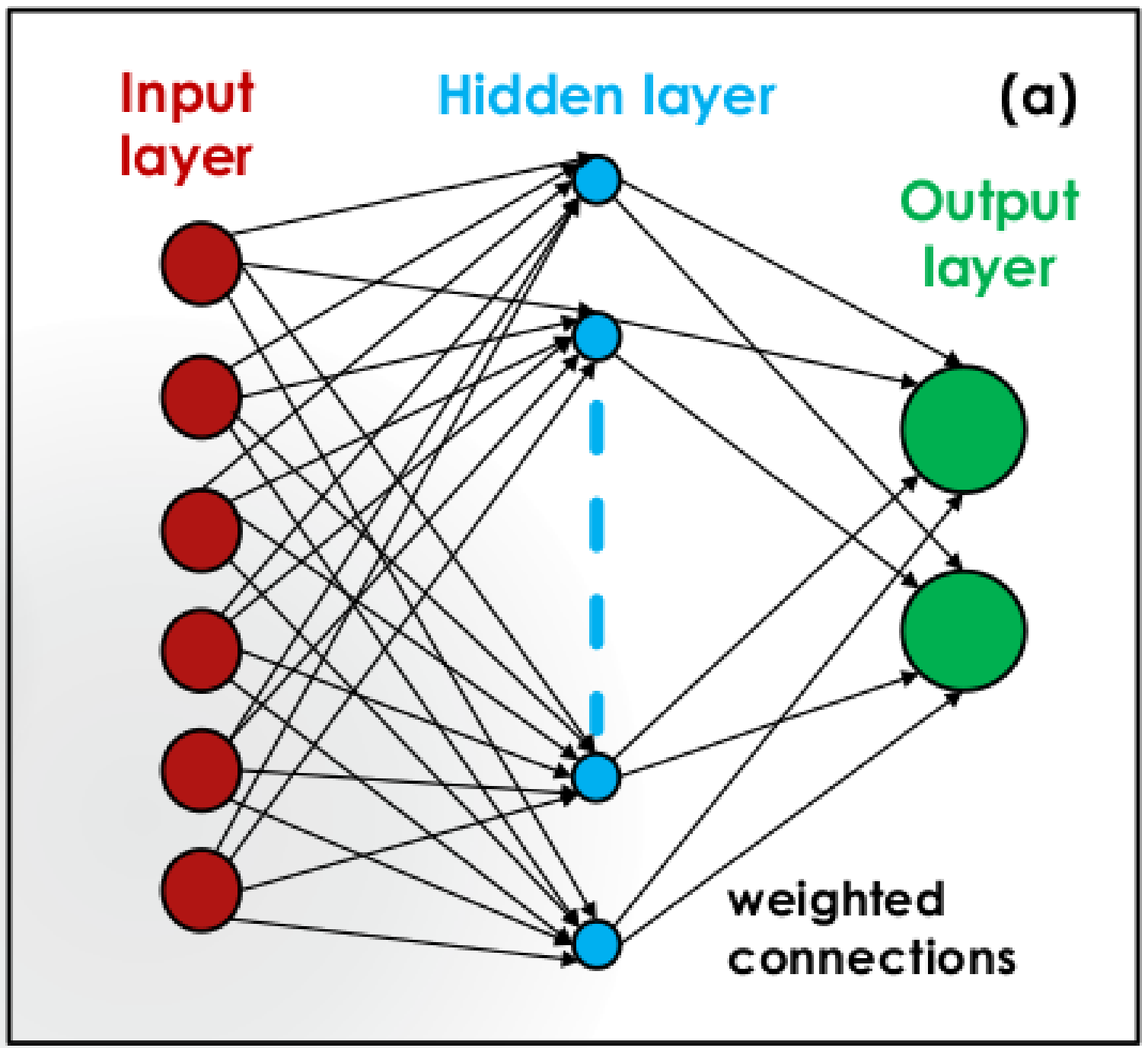}
	\includegraphics[width=0.30\textwidth]{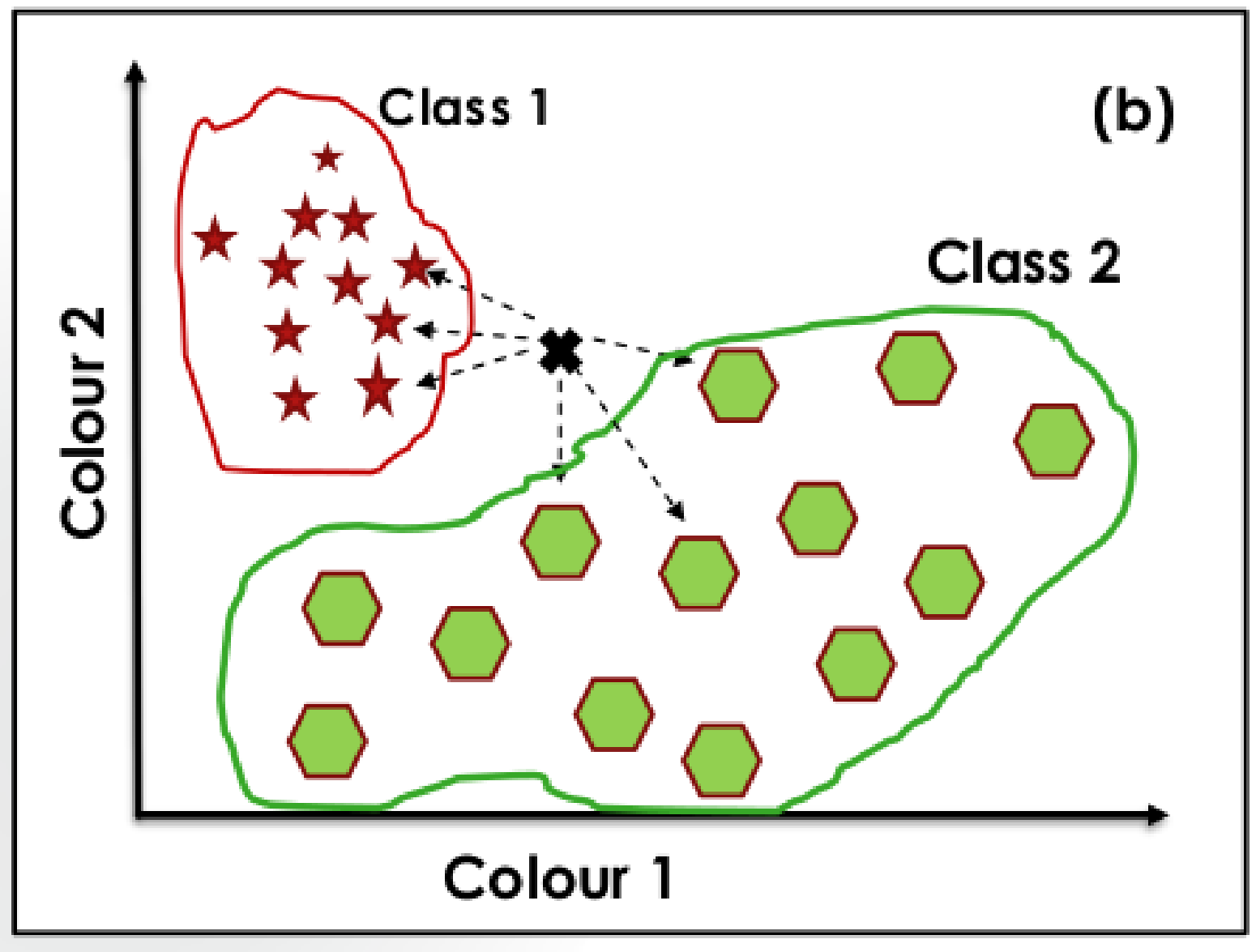}
	\includegraphics[width=0.30\textwidth]{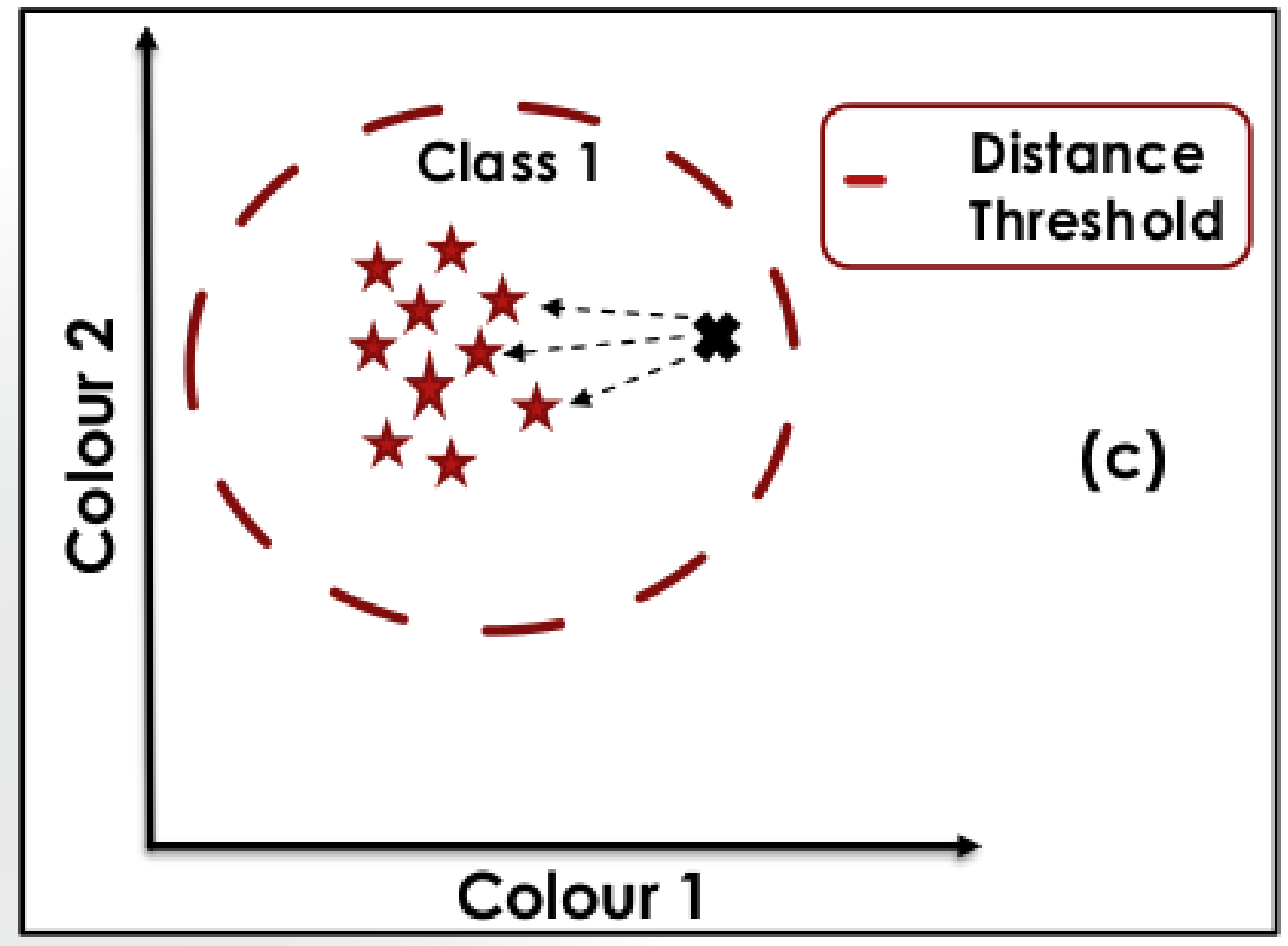}
	\caption{Schematic of the three classification techniques used in this study: (a) Neural networks, (b) \texttt{k}-NN Classification for \texttt{k} = 3, and (c) \texttt{k}-NN Threshold Distances for \texttt{k} = 3.}
  	\label{fig:Fig1}
\end{figure}

Neural networks are a family of machine learning algorithms developed for solving difficult pattern classification problems. Inspired by biological neural networks \citep{Beale1990}, they consist of individual processing units, called neurons or nodes. A network is constructed using layers of neurons. The first and last layer are the input and the output layer respectively, while the layers in between are designated as hidden layers. The first mathematical models of neurons, called perceptrons \citep{Rosenblatt1958}, were capable of identifying only linearly separable patterns. A network of such neurons formulated in a feed forward architecture with input, output and one or more hidden layers, called a feed forward neural network  (or a multilayer perceptron), was found to possess far greater classifying power. The layers are connected through one-directional weighted pathways between nodes in different layers. The input features for classification correspond to the nodes in the input layer, while the number of nodes in the the output layer is decided by the number of output classes.
The number of hidden neurons and layers are set according to the complexity of the classification problem, which decides the non-linearity of the network. A schematic of this method is shown in Fig.~\ref{fig:Fig1}(a).

Object classification was identified as one of the areas in astronomy where NeuN methods were likely to make an impact \citep{Miller1993}. What sets NeuN apart from the more conventional rule-based classifiers is their ability to learn from examples \citep{Cho1991}. This learnt information is stored in the network in the form of weights along the connecting pathways between individual neurons. This allows the network to generalise, making them capable of classifying patterns which may not be included in the initial training set. 

We create a simple feed-forward NeuN with the help of the MATLAB neural network toolbox. This employs supervised learning, with one hidden layer of neurons having 100 neurons. For training the network, we provide it with a training data set of example inputs and their corresponding desired outputs. Here, the inputs are a set of six colours (these are discussed in detail in Section~\ref{sec:colours}) for each object and the output is set to 0 or 1 depending on the class of the object. The network weights, which are initially set to random values, eventually store learnt patterns after training. During learning, the network weights are repeatedly adjusted in order to minimize the error between the obtained and expected output. This process, termed as backpropagation \citep{Rumelhart1986}, is repeated until either the entire training set is correctly classified, or the network is unable to minimize the error term further.


\subsection{\texttt{k}-Nearest Neighbour Approach to Classification (\texttt{k}-NN-C)}
In \texttt{k}-NN classification, the class of an object is decided based on its distance to a specific class of templates, where distances are defined in a multi-dimensional colour space. Usually, applications of the Nearest Neighbour methods consider the \texttt{k}-nearest templates (of any class) in a parameter space and decide the class based on the majority vote of these templates \citep{Popescu2018, Miettinen2018, Wallace2019, Akras2019}. However, in the present case we select \texttt{k} templates from each class and estimate an average distance to each class. We then determine the class to which the object has the shortest distance, and classify the object as belonging to that class.  
This approach can be considered as a modification of the threshold distance method used by \cite{Marengo2009}. While  these authors estimate the distance to a single class, we consider a training sample consisting of multiple classes: the brown dwarf and background. 
This approach is helpful for the current classification problem since the templates of one class are clustered in a small region and those of the other class (i.e. the background objects) are sparsely distributed. The same 6 colours used by the NeuN classifier as input parameters are again used here to create the multi-dimensional colour-colour space. 
A schematic of this method can be found in Fig.~ \ref{fig:Fig1}(b).  

An averaged Euclidean metric is preferred for multi-dimensional spaces as the distance does not increase with increase in the number of dimensions. While calculating the distance between the $i^{th} $ test object and the $j^{th} $ template in the training set for each colour $d_{l}(i,j) $, their photometric uncertainties (denoted by $\sigma_{i}$ and $\sigma_{j}$) are also taken into account. The \texttt{k}-NN distance of the $i$\textsuperscript{th} test sample to each class is the weighted average of the Euclidean distances $D(i,j) $ to the nearest \texttt{k} templates of that class, where the weights \textit{w(i, j)} are introduced to reduce the influence of isolated templates that happen to be much farther away than the nearest neighbors.  A Gaussian kernel is very effective for this task \citep{Marengo2009} which is given by Eqn.~\ref{(eq:wt)}. In addition to using uncertainties of the input colours in the \texttt{k}-NN distance equation, they also incorporate a sparseness factor $\sigma_{s} $, in order to account for lack of observed templates of a particular class in a given location in the colour-colour space. This factor is a measure of how far apart the templates are with respect to each input colour.

\begin{equation}
\sigma_s(j) =\frac{1}{k} \sqrt{\sum_{t=1}^{k} d_l(t,j)^2}
\label{sp}
\end{equation}

\begin{equation}
\sigma_l(i,j)=\sqrt{(\sigma_i^2 + \sigma_j^2)}+\sigma_s(j)
\label{eq:sl}
\end{equation}

\begin{equation}
w(i,j)=exp\left[-\left(\frac{D(i,j)}{k\times\sqrt{\sum_{l=1}^{N}\sigma_l(i,j)^2}}\right)^2\right] 
\label{(eq:wt)}
\end{equation}

The \texttt{k}-NN distance of the $i$\textsuperscript{th} test sample to a given class is given by the equation:
\begin{equation}
	D_{kNN}(i) =\frac{\sum_{j=1}^{k} D(i,j) \times w(i,j)}{\sum_{j=1}^{k} w(i,j)}.
	\label{eq:DkNN}
\end{equation}
 
The final classification is based on the minimum \texttt{k}-NN distance of the $i$\textsuperscript{th} test sample to each class. 
 
\subsection{\texttt{k}-NN Threshold Distances (\texttt{k}-NN-TD)}
In this method, the \texttt{k}-NN distance of each test object is calculated from a training sample consisting entirely of brown dwarfs, and objects whose \texttt{k}-NN distance is within a certain defined threshold, are classified as brown dwarfs. This can be visualised through the schematic given in Fig.~\ref{fig:Fig1}(c). This method is identical to the one used by \citet{Marengo2009} and the same formula, as given in Eqn.~(\ref{eq:DkNN}) is used for calculating the \texttt{k}-NN distance. This application requires templates only for the search class, relying on the assumption that the templates are an accurate representation of the class, and that the selected features (colours) chosen in the analysis are sufficient to provide an effective discrimination. 

The threshold distance, denoted by $D_{th}$, and the number of neighbors, \texttt{k}, are optimized for maximum completeness and rejection efficiency using the bootstrap method \citep{Hastie2001}. The  completeness ($\mathcal{C}$) and rejection ($\mathcal{R}$) efficiencies are defined as follows:

\begin{equation}
	\text{Completeness,\,\,} \mathcal{C}= \frac{\text{No. of true positives}}{\text{No. of expected brown dwarfs}}
\end{equation}
\begin{equation}
	\text{Rejection Efficiency,\,\,} \mathcal{R}= \frac{\text{No. of true rejections}}{\text{Total no. of background objects}}
\end{equation}

Samples of the template and background classes are created and then tested for $\mathcal{C}$ and $\mathcal{R}$ for different values of \texttt{k} and $D_{th}$. In order to standardize this approach, we select the \texttt{k} and $D_{th}$ values which give the highest product of $\mathcal{C}$ and $\mathcal{R}$.

\subsection{Ensemble Classifier}
We use the parametric (NeuN) as well as the two non-parametric (\texttt{k}-NN) techniques to create an ensemble classifier. Such a classifier will identify brown dwarfs by factoring in the outputs from all the individual classifiers and making the final decision on the basis of a majority vote. In other words, if either two or all three methods classify an object as a brown dwarf candidate, then the ensemble method would label this object as a brown dwarf candidate and vice-versa. In this method, each individual classifier is given equal weightage. This helps to reduce the misclassification due to the shortcomings of any one classifier. We attempt to quantify the performance of the ensemble classifier by calculating $\mathcal{C}$ and $\mathcal{R}$ in each case.

\section{Features and Data for Classification}
\label{sec:Sec3}
The dearth of work done in this type of automated brown dwarf classification problem prompted us to create training datasets from scratch and we have decided on a set of features to be used for classification, after going through a wide variety of pre-existing literature. These are described in this section.

\subsection{Colours used as input features}
\label{sec:colours}

 The data used for photometric classification includes the brightness of objects as measured in WISE and the Two Micron All-Sky Survey \citep[2MASS]{Strutskie2006}. The WISE bands W1, W2 and W3 have been incorporated in the analysis. W4 has been excluded as the angular resolution in this band is lower than the other bands by a factor of $\sim2$. In addition, the photometric uncertainties of many objects are not available in W4. Moreover, we find that its inclusion did not show any significant improvement in the results. In 2MASS, all the three bands: i.e. J (1.25 $ \mu $m), H (1.65 $ \mu $m), K$_{\rm s}$ (2.17 $ \mu $m) bands have been considered. With these bands, it is possible to identify the near-infrared spectral features associated with H\textsubscript{2}O in the atmospheres of the L and T dwarfs \citep{Stephens2004}. We have selected six colours  based on the spectral characteristics of brown dwarfs in the WISE and 2MASS filter combinations \citep{Kirkpatrick2011,Marengo2009,Faherty2016, Zhang2018}. The colours used in this work are: W1-W2, W2-W3, J-H, J-W1, J-W2, J-K$_{\rm s}$, and the spectral characteristics of brown dwarfs that they describe are listed in Table~\ref{table:colours}. 


\begin{table}
	\centering
	\caption{Photometric colours used as features for brown dwarf classification, using WISE and 2MASS filters.}
	\begin{tabular}{|c|c|}
		\hline \hline
		Colour		   & Characteristic           \\ \hline
		W1-W2          & Methane absorption in W1          \\
		W2-W3          & Methane absorption in lower bands \\
		J-H            & H\textsubscript{2}O absorption in J               \\
		J-W1           & H\textsubscript{2}O absorption in J               \\
		J-W2           & H\textsubscript{2}O absorption in J               \\
		J-K$_{\rm s}$  & Presence of methane               \\ \hline \hline
	\end{tabular}
	\label{table:colours}
\end{table}


\subsection{Training Samples}
The training sets are constructed using different combinations of brown dwarfs and background objects. The selection of templates used for the classification is described below.

\begin{enumerate}
	\item Brown dwarfs - The brown dwarfs used as templates for this work have been taken based on availability of infrared data in all the required bands. For this we have used brown dwarfs from the following research papers: \citet{Kirkpatrick2011}, \citet{Mace2013}, and \citet{Best2018}.\\
	\item Background objects - In this classification problem, background objects include other sources such as stars in different evolutionary stages or galaxies. For this work, we have attempted to incorporate a variety of background object templates and only those have been included in the training sets, which were felt to have an effect on the classification on brown dwarfs and, hence, were likely contaminants. The objects were taken from existing literature and include NLS1 galaxies \citep{Chen2017A}, Ap and Am stars \citep{Chen2017B}, Young Stellar Objects \citep[YSOs;][]{Su2014,Fischer2016}, Red Giants \citep{Anders2017}, K-type stars \citep{Pecaut2016}, and M- dwarfs \citep{Best2018}. Along with all these known background objects, some objects have been taken randomly from the WISE All-Sky point source catalog \citep{Wright2010} by applying magnitude cuts so that they are brighter than the known brown dwarf templates in each band. 
\end{enumerate} 

Apart from the above objects, some artificial samples are created for both classes with the characteristics of the known templates. This is carried out in a manner similar to the bootstrap implementation described by \citet{Marengo2009}.


\section{Optimal Training Sets and Performance of Classification Methods}
\label{sec:Sec4}
It is well known that for any classification problem, diverse training data sets are of great importance. But, when it comes to brown dwarfs, few attempts have been made in this regard. Here, we have experimented with different catalogs of known brown dwarfs as well as various other stars and galaxies to create the brown dwarf and background classes, respectively. Distinct training sets were generated by combining specific template groups. 

For each set, $\mathcal{C}$ and $\mathcal{R}$ are employed as the validation metrics. Each set is randomly divided into three parts, in the ratio 70:15:15, and labelled as training, validation and test samples, respectively. The validation set is used to tune the parameters of a classifier, for example, to tune the weights in a neural network. In the \texttt{k}-NN-TD method, the validation set was used to fix the distance threshold value which gave the maximum  $\mathcal{C}$- $\mathcal{R}$ efficiency product. The test set is then used to assess the performance of the final tuned classifier by the estimation of $\mathcal{C}$and $\mathcal{R}$. We carried out a rigorous 5-fold cross-validation for the verification of these efficiencies. In this method, each dataset is divided randomly into 5 parts. One part is taken as the test set whereas the other 4 are combined to form a training set. This testing is repeated for all the 5 parts, and the average $\mathcal{C}$ and $\mathcal{R}$ evaluated. We find that the efficiencies computed by both these methods are very similar (match within $\sim 90$\%). 

Based on the values of $\mathcal{C}$ and $\mathcal{R}$, the composition of the sets were modified to select the best training sets for this particular problem. These are described below.


\begin{table*}
	\centering
	\caption{Composition of the 2-class Training Sets.}
	\begin{tabular}{c|c|c|c|c|c}
		\hline\hline
		Set & \multicolumn{2}{c|}{Brown Dwarfs} & \multicolumn{2}{c|}{Background Objects} & \\ \hline
		& Composition & \begin{tabular}[c]{@{}c@{}}No. of\\ objects\end{tabular} & Composition & \begin{tabular}[c]{@{}c@{}}No of \\ objects\end{tabular} & \begin{tabular}[c]{@{}c@{}}Total no.\\ of objects\end{tabular} \\ \hline
		A & \begin{tabular}[c]{@{}c@{}}Kirkpatrick,Thompson,\\ Simulated\end{tabular} & 1430 & \begin{tabular}[c]{@{}c@{}}WISE background objects\\  (Real\&Simulated), NLS1 galaxies\end{tabular} & 1200 & 2630 \\
		B & \begin{tabular}[c]{@{}c@{}}Kirkpatrick,Thompson,\\ Simulated\end{tabular} & 1430 & \begin{tabular}[c]{@{}c@{}}WISE background objects (Real\&Simulated),\\ NLS1 galaxies, Fischer YSOs\end{tabular} & 1275 & 2705 \\
		C & \begin{tabular}[c]{@{}c@{}}Kirkpatrick,Thompson,\\ Best T-dwarfs\end{tabular} & 344 & \begin{tabular}[c]{@{}c@{}}NLS1 galaxies, Fischer YSOs, \\ Ap \& Am stars, Red Giants, K-type stars\end{tabular} & 325 & 669 \\ \hline\hline
	\end{tabular}
	\label{table:2class1}
\end{table*}



\begin{figure*}
	\includegraphics[width=0.30\textwidth]{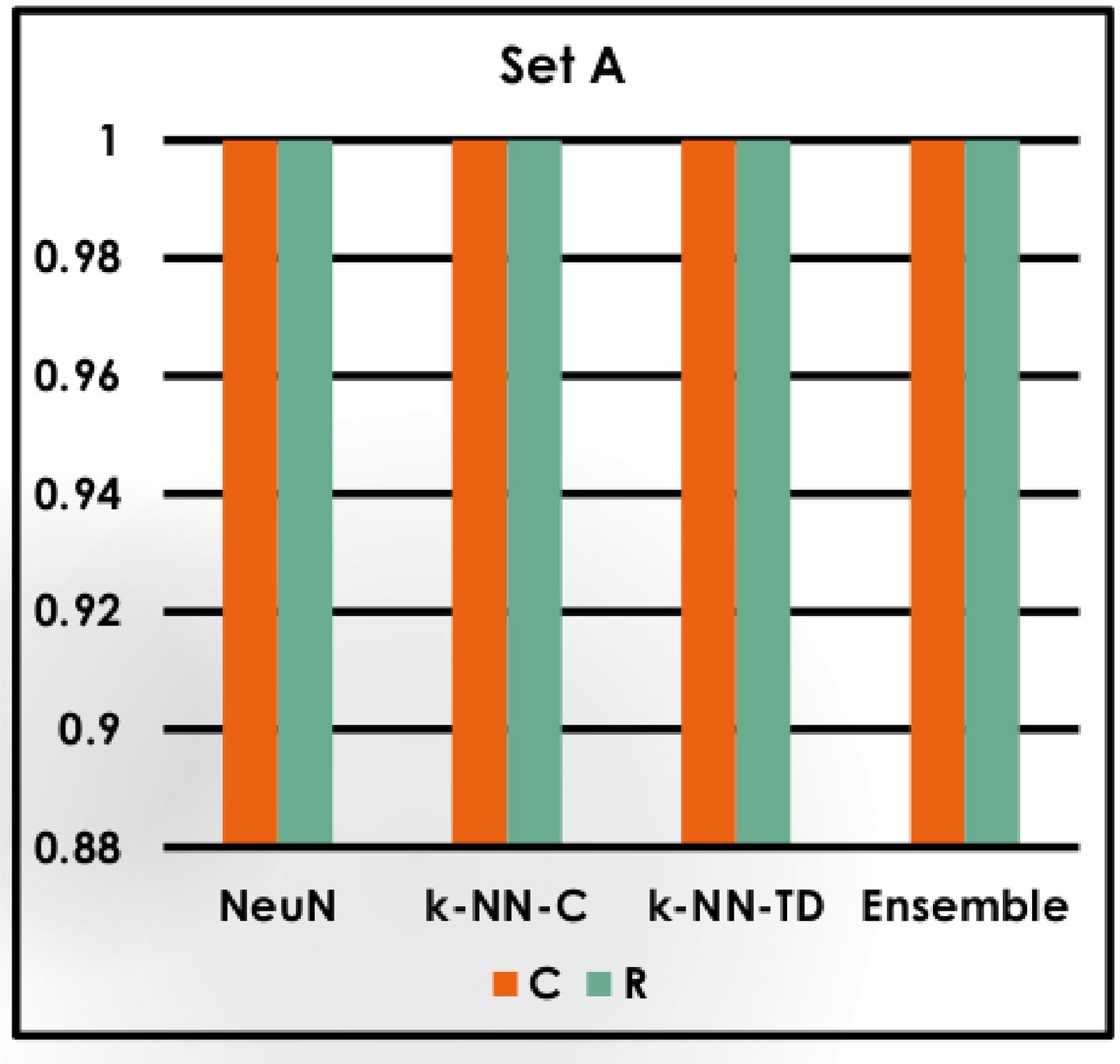}
	\includegraphics[width=0.30\textwidth]{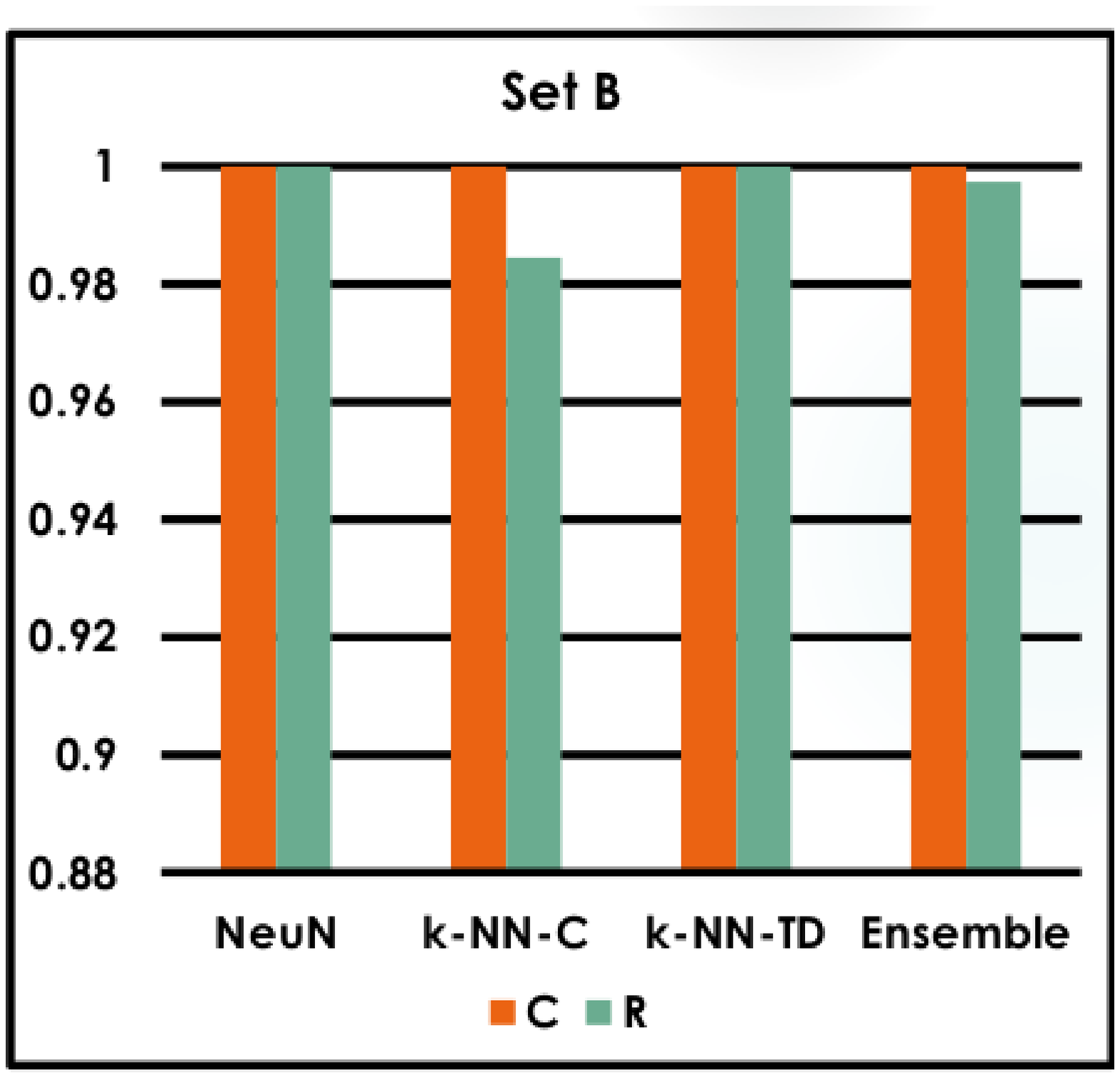}
	\includegraphics[width=0.30\textwidth]{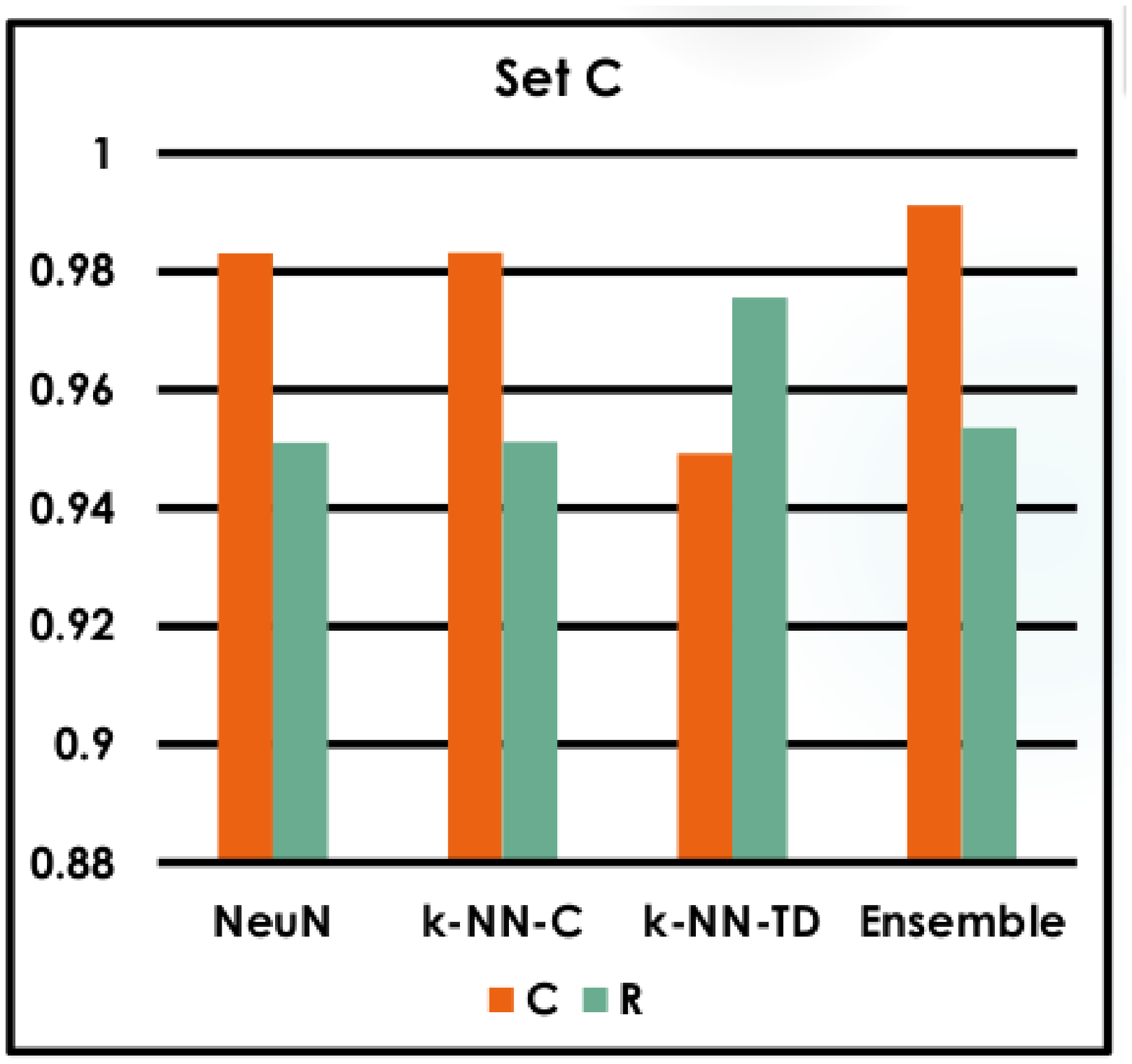}
	\caption{Completeness and Rejection Efficiencies of the 2-class training sets A, B and C, described in Sec~\ref{sec:Sec4}.}
	\label{fig:Fig2}
\end{figure*}


\subsection{2-Class Training Sets}

The 2-class training sets include the classes of (i) brown dwarfs, and (ii) background objects. Three distinct groups were constructed for this purpose, using the combinations of objects described below. All the classification methods were applied on the training sets and the results are presented. 

The \texttt{k}-NN-TD method is also discussed here although it employs only one class of brown dwarf templates as its training set. The threshold distance effectively demarcates the feature space into two classes, with all the objects falling within the threshold considered as brown dwarf candidates and the rest taken as other (background) objects. Thus, it is equivalent to a 2-class classification technique in this study, and we have compared its performance with the rest of the 2-class methods. We note that for all the 2-class training sets described below, only the brown dwarf template class in each group is used by the \texttt{k}-NN-TD classification method.

\begin{enumerate}
	\item Set A - This set comprises of the brown dwarfs from the sample of \citet{Kirkpatrick2011} and \citet{Mace2013}, in addition to the simulated objects created from these samples. The background class includes random WISE background objects and the corresponding simulated objects alongwith NLS1 galaxies. Note that the NLS1 galaxy sample also includes point-like sources, with colours similar to those of brown dwarfs in the bands under consideration. The resultant efficiency values, on application of the classification techniques, are shown in Fig.~\ref{fig:Fig2}. We find that the efficiencies are 100\%. This is due to the relatively small variety of objects in the training samples, and a marked difference between the characteristic colours of the two classes, i.e., brown dwarfs and the WISE background.
	\item Set B - The composition of this set is similar to Set A, the difference being the addition of YSO templates to the background class \citep{Fischer2016}. The inclusion of YSOs lowers the $\mathcal{R}$ of the \texttt{k}-NN-C method to $\sim98$\%, which in turn lowers the $\mathcal{R}$ of the ensemble classifier by $\sim0.3$\% (see Fig.~\ref{fig:Fig2}). This could be because of the fact that YSOs have colours similar to those of brown dwarfs in the infrared bands with few being misclassified. We note that $\mathcal{C}$ values remain unaffected.
	\item Set C - This set includes the brown dwarfs from the samples of \citet{Kirkpatrick2011}, \citet{Mace2013} and \citet{Best2018}. The background class comprises of all known objects (NLS1 galaxies, Ap and Am stars, YSOs, Red Giants, K-type stars, and M- dwarfs). The resultant efficiencies display a lowering of the $\mathcal{C}$ and $\mathcal{R}$ values with respect to sets A and B, see Fig.~\ref{fig:Fig2}. The ensemble classifier performs marginally better than the others. The NeuN and \texttt{k}-NN-C methods give similar values of $\mathcal{C}$~(98\%) and $\mathcal{R}$~(95\%). \texttt{k}-NN-TD returns a better $\mathcal{R}$~(98\%) but a relatively lower $\mathcal{C}$~(95\%). It would appear that an increase in the variety of the samples has affected the efficiencies. Such diversification of training set is expected to help improve the $\mathcal{R}$ of the classifiers during testing on a real scenario, as more non-brown dwarf templates start finding representation in the background class. 
\end{enumerate}

The composition of the different training sets are summarised in Table \ref{table:2class1}. 


\begin{table*}
	\centering
	\caption{Efficiencies of 2-Class Training Sets, described in Sec~\ref{sec:Sec4}..}
\begin{tabular}{crrrrrrrr}
\hline
\multicolumn{1}{c}{}    & \multicolumn{4}{c}{Completeness}                                                                                                                                                                                                                                                                                 & \multicolumn{4}{c}{Rejection Efficiency}                                                                                                                                                                                                                                                                       \\ \hline
\multicolumn{1}{c}{Set} & \multicolumn{1}{r}{\begin{tabular}[r]{@{}r@{}}NeuN\\ (\%)\end{tabular}} & \multicolumn{1}{r}{\begin{tabular}[r]{@{}r@{}}\texttt{k}-NN-C\\ (\%)\end{tabular}} & \multicolumn{1}{r}{\begin{tabular}[r]{@{}r@{}}\texttt{k}-NN-TD\\  (\%)\end{tabular}} & \multicolumn{1}{r}{\begin{tabular}[r]{@{}r@{}}Ensemble \\ (\%)\end{tabular}} & \multicolumn{1}{r}{\begin{tabular}[r]{@{}r@{}}NeuN\\ (\%)\end{tabular}} & \multicolumn{1}{r}{\begin{tabular}[r]{@{}r@{}}\texttt{k}-NN-C\\ (\%)\end{tabular}} & \multicolumn{1}{r}{\begin{tabular}[r]{@{}r@{}}\texttt{k}-NN-TD\\ (\%)\end{tabular}} & \multicolumn{1}{r}{\begin{tabular}[r]{@{}r@{}}Ensemble\\ (\%)\end{tabular}} \\ \hline
A                       & 100                                                                     & 100                                                                       & 100                                                                         & 100                                                                          & 100                                                                     & 100                                                                       & 100                                                                        & 100                                                                         \\
B                       & 100                                                                     & 100                                                                       & 100                                                                         & 100                                                                          & 100                                                                     & 98.5                                                                     & 100                                                                        & 99.7                                                                       \\
C                       & 98.3                                                                   & 98.3                                                                     & 94.9                                                                       & 99.1                                                                        & 95.1                                                                   & 95.1                                                                     & 97.6                                                                      & 95.35                                                                       \\ \hline
\end{tabular}
	\label{table:2class2}
\end{table*}


\subsection{3-Class Classification}
\label{sec:3class}
The spectral categories of brown dwarfs (L, T, Y) can be utilized to make finer constraints in the feature space, for their identification. The L category of objects cannot be strictly considered as a brown dwarf class as it can also include low-mass stars. Therefore, in order to better evaluate the effects of L-dwarfs in the training sets on the classification of both types of objects (brown dwarfs and background), the NeuN and \texttt{k}-NN-C classifiers have been built with 3 categories: (i) T- \& Y-dwarfs, (ii) L-dwarfs, and (iii) background objects. Each test object is classified into one of the three classes based on the features. The training set created for the 3-class classification contains 2698 objects. There are  335 objects in the first category which has T- and Y-dwarfs taken from \citet{Kirkpatrick2011} and \citet{Mace2013}, along with T-dwarfs from \citet{Best2018}. There are 1087 objects taken from the above samples in the second category (i.e. L-dwarfs). There are 1275 background objects comprising of random WISE background, NLS1 galaxies, YSOs from the \citet{Fischer2016} catalog and M-dwarfs. This composition is listed in Table \ref{table:3class1}. For this training set, only \texttt{k}-NN-C and NeuN methods are applicable as they cater to multiple classes. 

\begin{table}
	\centering
	\caption{3 Class Classification-Training Set Composition}
	\begin{tabular}{ccc}
		\hline
		Set & Composition & \begin{tabular}[c]{@{}c@{}}No. of \\ objects\end{tabular} \\ \hline
		3A & \begin{tabular}[c]{@{}c@{}}Best Dwarfs(1000 M,L,T), \\ Kirkpatrick+Thompson,\\ WISE background (Real),\\  NLS1 galaxies, CM YSOs\end{tabular} & \begin{tabular}[c]{@{}c@{}}2698\\ (Bg-1275, \\ L-1087,\\  T-335)\end{tabular} \\ \hline
	\end{tabular}
	\label{table:3class1}
\end{table}



\begin{figure*}
	\includegraphics[width=0.30\textwidth]{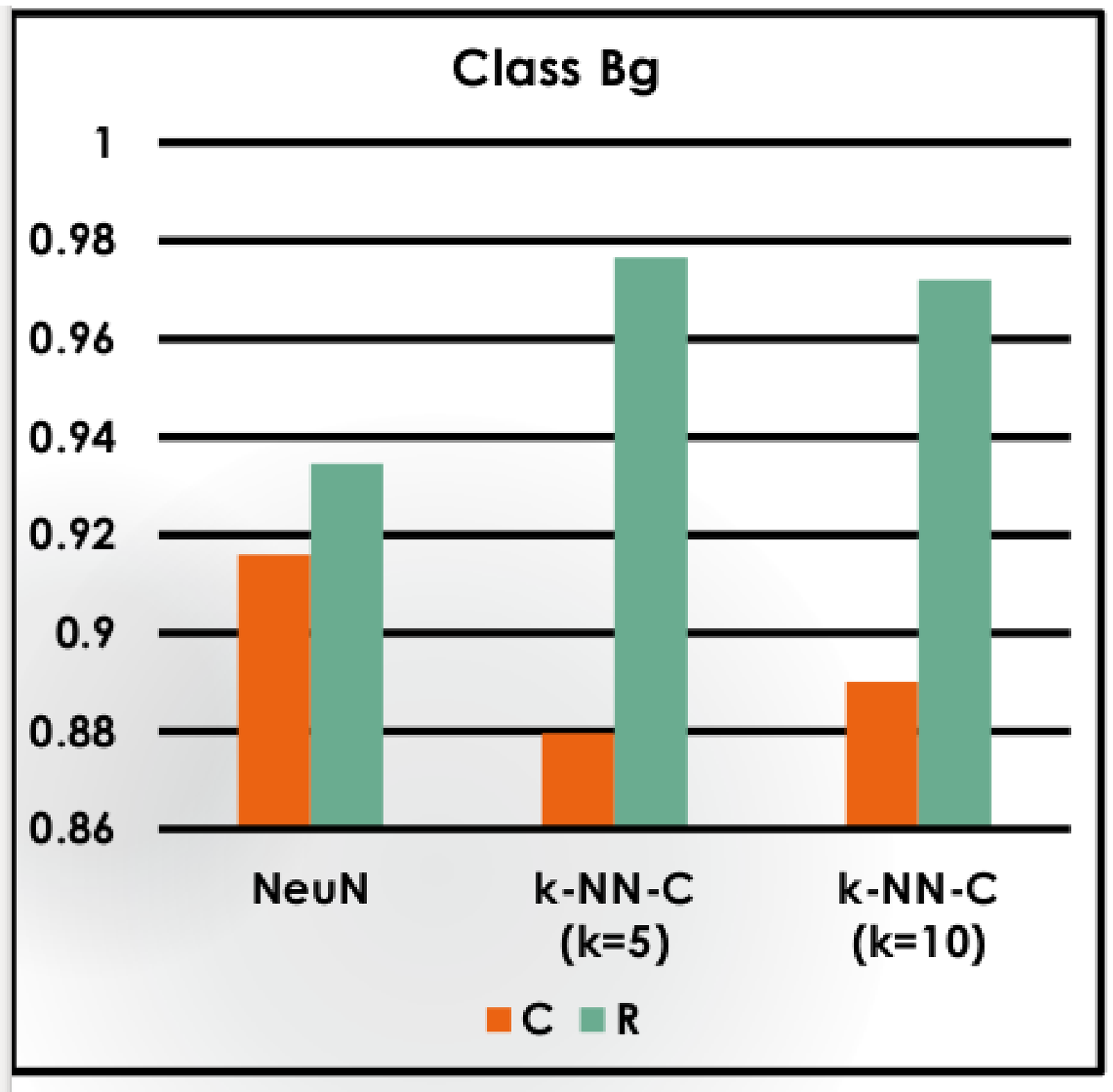}
	\includegraphics[width=0.30\textwidth]{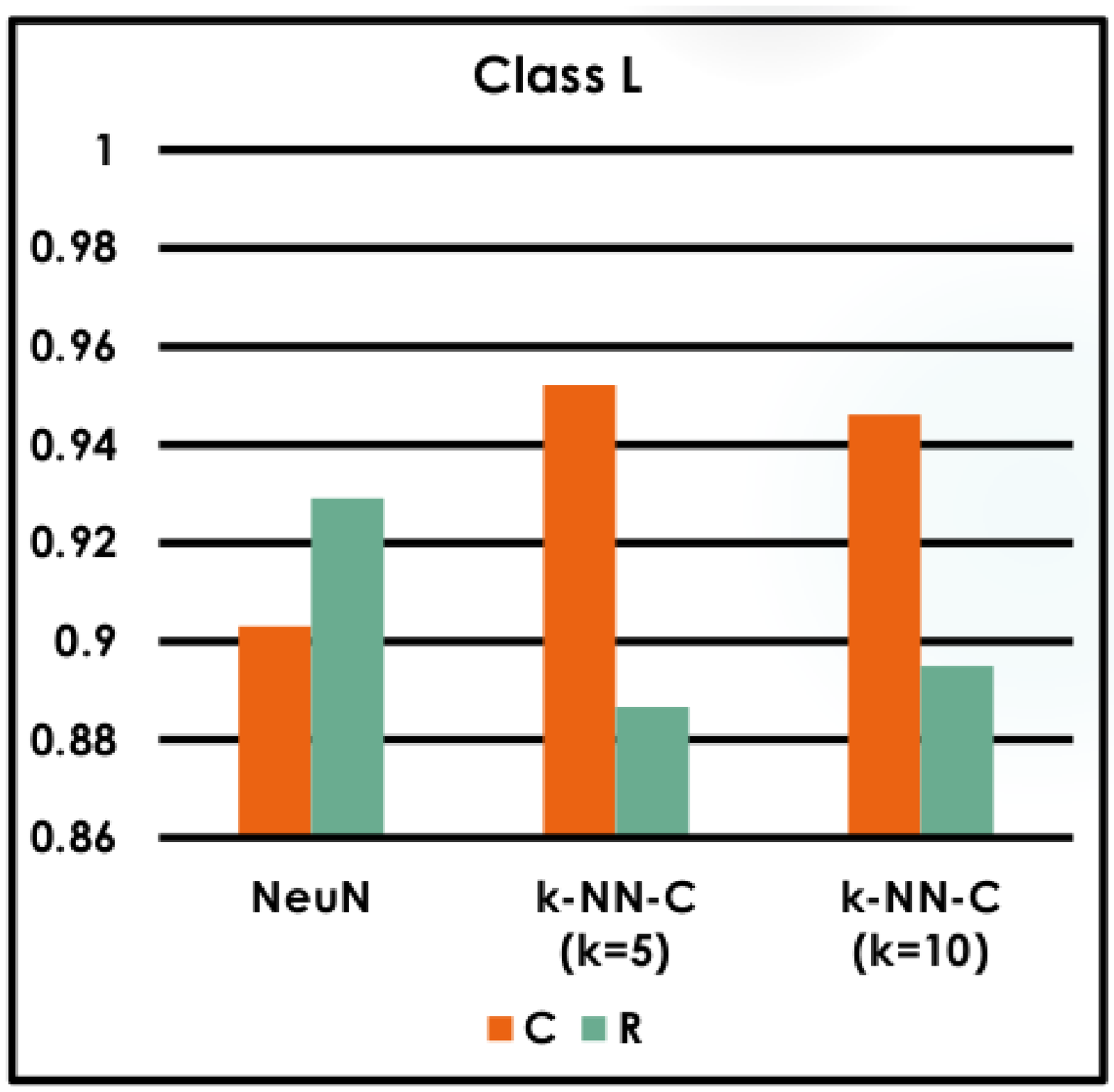}
	\includegraphics[width=0.30\textwidth]{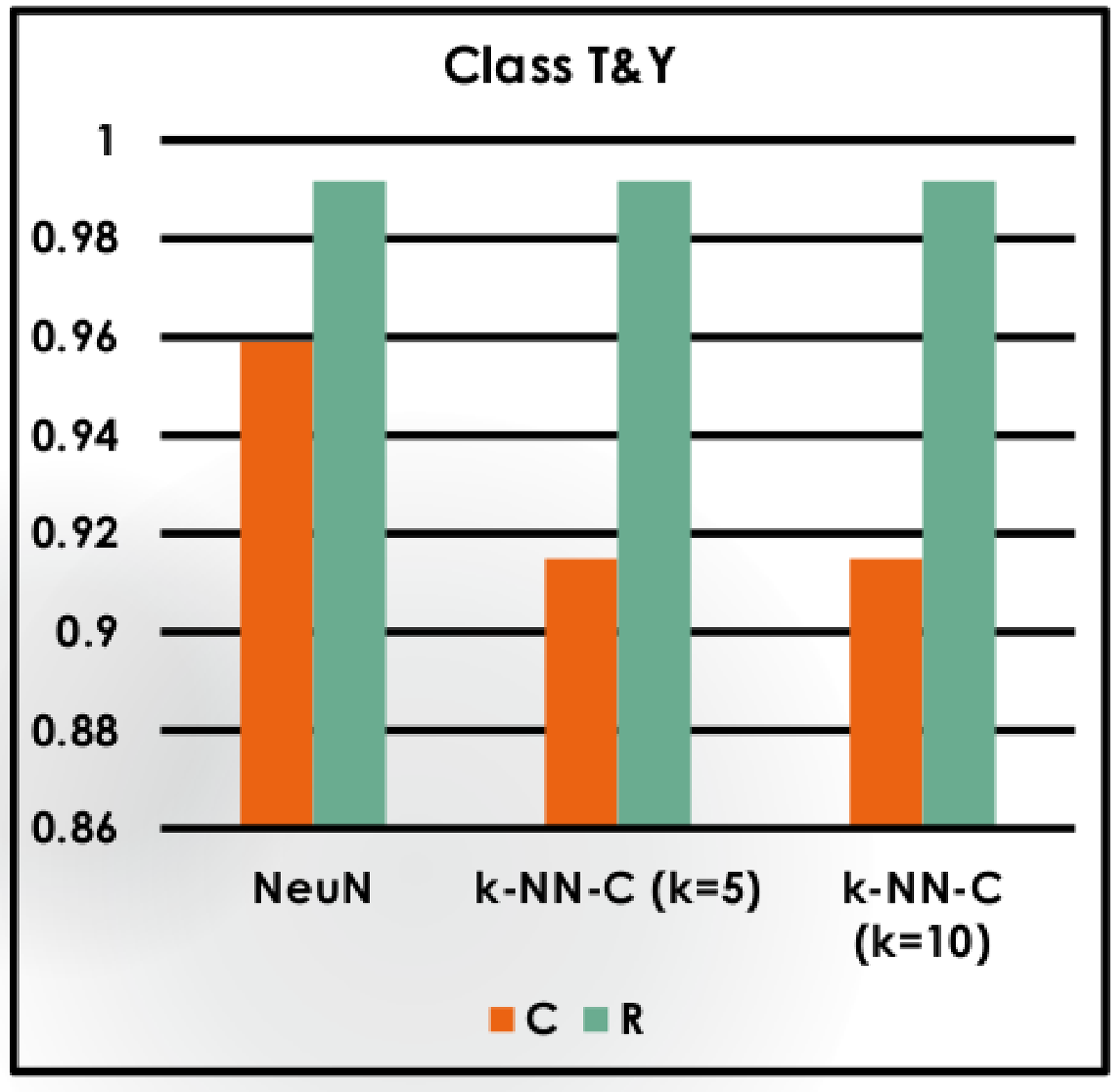}
	\caption{Completeness and Rejection Efficiencies of the 3-Class Classification (Background denoted by Bg, L-dwarfs and T, Y-dwarfs), described in Sec~\ref{sec:3class}.}
	\label{fig:Fig3}
\end{figure*}



\begin{figure*}
	\includegraphics[width=0.40\textwidth]{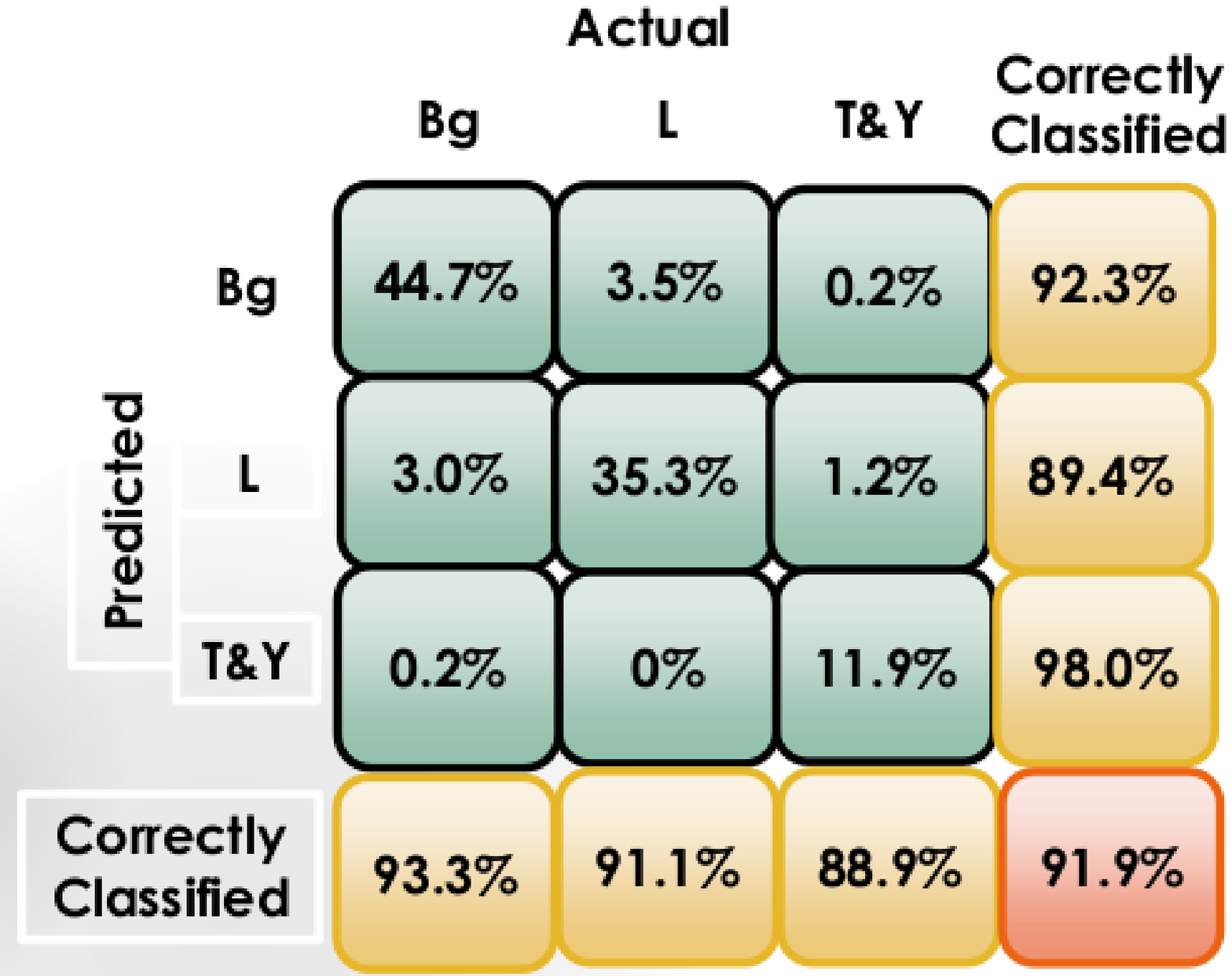}
	\includegraphics[width=0.40\textwidth]{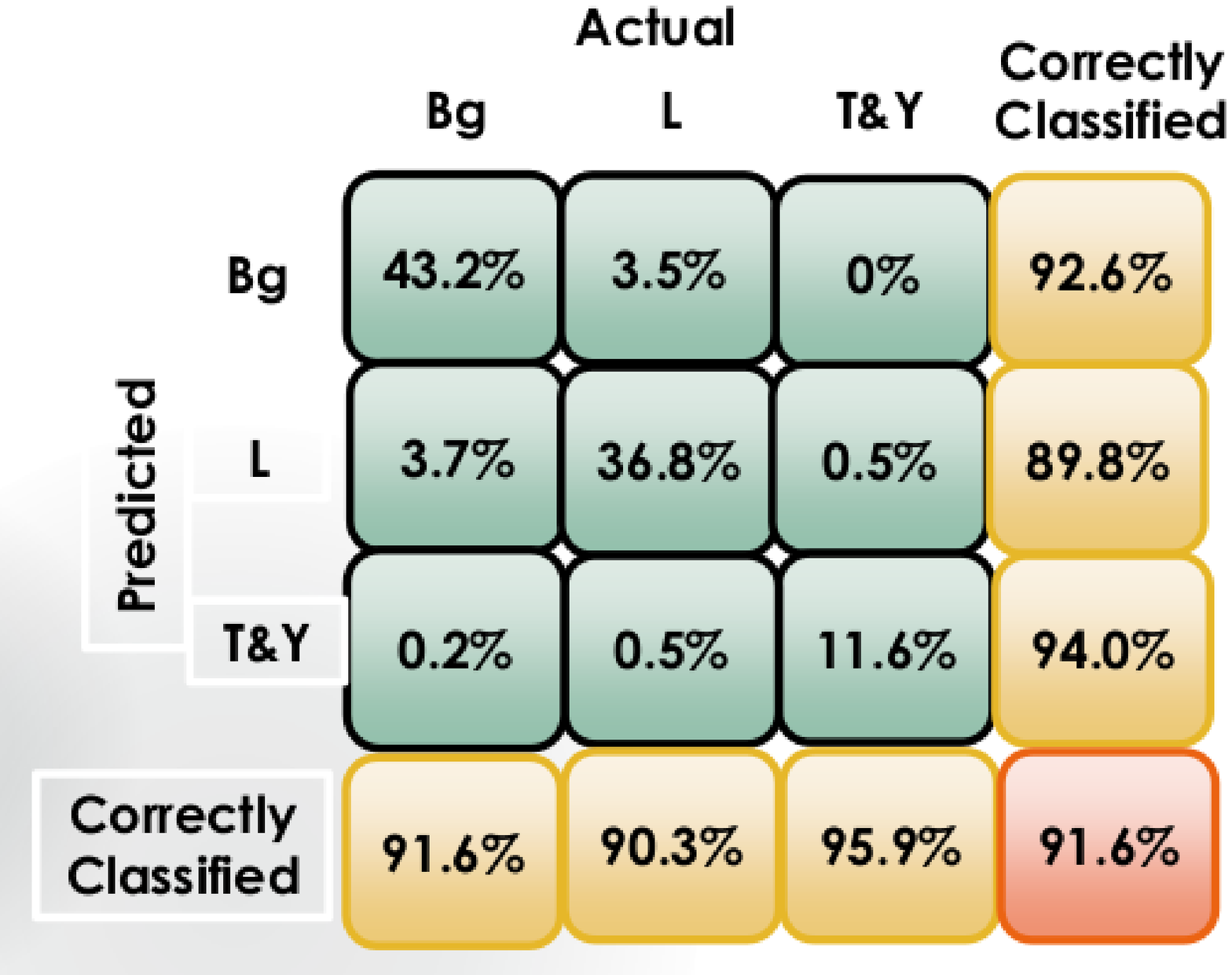}
	\caption{Confusion matrix for the 3-Class Classification (Background denoted by Bg, L-dwarfs and T, Y-dwarfs) by NeuN classifier for  (left) validation and (right) test sets.}
	\label{fig:Fig4}
\end{figure*}



\begin{table*}
	\centering
	\caption{Efficiencies of 3-Class Training Set.}
	\begin{tabular}{ccrrrrrr}
		\hline\hline
		&  & \multicolumn{3}{c}{Completeness} & \multicolumn{3}{c}{Rejection Efficiency} \\ \hline
		Set & \begin{tabular}[c]{@{}c@{}}Output\\ Class\end{tabular} & \begin{tabular}[r]{@{}r@{}}NeuN\\ (\%)\end{tabular} & \begin{tabular}[r]{@{}r@{}}\texttt{k}-NN-C\\ k=5 (\%)\end{tabular} & \begin{tabular}[r]{@{}r@{}}\texttt{k}-NN-C\\ k=10 (\%)\end{tabular} & \begin{tabular}[r]{@{}r@{}}NeuN\\ (\%)\end{tabular} & \begin{tabular}[r]{@{}r@{}}\texttt{k}-NN-C\\ k=5 (\%)\end{tabular} & \begin{tabular}[r]{@{}r@{}}\texttt{k}-NN-C\\ k=10 (\%)\end{tabular} \\ \hline
		& Bg & 93.3 & 87.0 & 87.0 & 92.9 & 97.6 & 96.6 \\
		Validation & L & 91.1 & 94.3 & 94.3 & 93.1 & 90.7 & 90.3 \\
		& T and Y & 88.9 & 95.8 & 95.8 & 99.7 & 97.5 & 98.3 \\ \hline
		& Bg & 91.6 & 88.0 & 89.0 & 93.5 & 97.7 & 97.2 \\ 
		Test & L & 90.3 & 95.2 & 94.6 & 92.9 & 88.7 & 89.5 \\
		& T and Y & 95.9 & 91.5& 91.5 & 99.2 & 99.2 & 99.2 \\ \hline \hline
	\end{tabular}
	\label{table:3class2}
\end{table*}


The resultant efficiencies of this classification are displayed in Table \ref{table:3class2} and Fig.~\ref{fig:Fig3}. We observe that $\mathcal{R}$ values for the T- and Y-dwarfs are high ($>97$\%), but the $\mathcal{C}$ values are relatively lower ($\mathcal{R}\ge88$\%). This can be attributed to the misclassification of early T-dwarfs as L-dwarfs. The L-dwarfs have lower $\mathcal{R}$ values ($\sim 91\%$) as compared to the T- and Y-dwarfs but the $\mathcal{C}$ values are comparable ($\ge 90\%$).The background class exhibits a trend similar to the T- and Y-dwarfs: low $\mathcal{C}$ values($\ge87$\%) but high $\mathcal{C}$ values ($\ge87$\%).

The confusion matrix for the 3-class NeuN classifier is illustrated in Fig.~\ref{fig:Fig4}. A confusion matrix is used to describe the performance of a classifier with multiple classes on a set of test data for which the true values are known. The rows of the matrix correspond to the class predicted by the NeuN classifier for a test object while the columns correspond to the actual class of the object. The cells along the diagonal represent the fraction of test objects correctly classified by the network. From the figure, we see that a relatively larger fraction of L-dwarfs (3.5\%) are being misclassified as background when compared to T- \& Y-dwarfs ($\sim$ 0.5\% or lower). The fraction of correctly classified test objects of that class, given by the last cell at the end of each row, is lower for the L-dwarfs when compared to the T- and Y-dwarfs, as expected. All these results imply that a small fraction of the L-dwarfs are closer to the background templates, than brown dwarfs. The lowermost cell towards the right end gives the $\mathcal{C}$ of the network for the given sample. We find that the $\mathcal{C} > 91.5$\% for NeuN. It is to be noted here that the ensemble classifier cannot be applied for the 3-class training sets as there are only two methods, and hence a majority vote would not be possible in some cases.  

\subsection{Performance of Classification Methods}
We summarise the performance of the classification techniques here. We find that all the three classification methods perform well on the training sets considered, with $\mathcal{C}\ge87$\% and $\mathcal{R}\ge88$\%. In the 2-class classification, the ensemble classifier performs as well as NeuN for the sets A and B, and marginally outperforms the other classifiers for set C. In the classification methods using 3-class training sets, it would appear that both NeuN and \texttt{k}-NN-C perform equally well. Thus, we see that the methods outlined above perform unequivocally well on the diverse training sets that have been created. 


\section{Testing on Regions in Space}
\label{sec:Sec5}
Having analysed the performance of the classification methods to known objects, we apply them to objects in three different regions of the sky. Two of them, Hercules and Serpens, have been selected due to the relatively larger number density of known brown dwarfs in these regions. The third arbitrary region was selected based on the attribute that it should not have any known brown dwarf. For this, we considered a region towards the Lyra. The details of these three regions are given below:
\begin{enumerate}
	\item \textit{A part of the constellation Serpens}:  This comprises a region of size 4\degr$\times$9\degr, centred on RA (J2000)=246.500\degr, Dec (J2000)=4.500\degr.
	\item \textit{The constellation Hercules}: This comprises a region of size 2\degr$\times$2\degr, centred on RA  (J2000)=268.000\degr, Dec  (J2000)=17.000\degr.
	\item \textit{The constellation Lyra}: This comprises a region of size 2\degr$\times$2\degr  centred on RA (J2000)=275.410\degr, Dec (J2000)=32.450\degr.
\end{enumerate}

We include all sources for analysis in a given region whose infrared photometry in the requisite bands are available in the AllWISE Source Catalog \citep{Cutri2013}. The catalog also lists the association of sources with 2MASS Point and Extended Source Catalog. The WISE photometric magnitudes, their uncertainities and the associated  2MASS photometry and uncertainities have been used for the classification. We restrict the search to point sources by taking only those sources with 2MASS \texttt{extended source flag} = 0, which indicates that the morphology of the source is consistent with a point source.

The Hercules region has 3 known T-dwarfs, one identified by \citet{Mace2013} and two by \citet{Best2018}. The Serpens region has 5 T-dwarfs, all of them identified by \citet{Best2018}, and two of them were identified by \citet{Mace2013}. Therefore, if a classifier is able to identify the known brown dwarfs in these regions, it would further validate the performance of the classification techniques. For these regions, it is not possible to calculate $\mathcal{R}$ as we do not know the exact number of brown dwarfs and background objects that are present. Therefore, in order to quantify how effectively the classifier rejects background objects, we define a new parameter called the rejection ratio (\textit{RR}) in the following way. 

\begin{equation}
	\text{Rejection Ratio,\,\,} \mathit{RR}= \frac{\text{No. of objects rejected}}{\text{Total no. of objects in the region}}
\end{equation}

\subsection{Serpens}
The Serpens region contains 21,022 objects with photometry in all the requisite bands. This region contains 5 T-dwarfs. The classifier techniques developed till now are applied to the region. We find that, among the 2-class methods, NeuN performs exceptionally well and identifies all the five known brown dwarfs with \textit{RR} of 99.9\% and above, see Table~\ref{table:Regions2class}. In the \texttt{k}-NN-C method, all the training sets, except Set A for k=10, identify all known dwarfs with a reasonably high \textit{RR}. The performance of the \texttt{k}-NN-TD method, is quite poor with sets A and B unable to identify any known dwarf, whereas set C identifies all known dwarfs, albeit with a lower \textit{RR} than the other methods (\textit{RR} $\sim$ 94\%). The ensemble classifier also works well with sets B and C identifying all known dwarfs and set A identifying 4 out of 5. The \textit{RR} values for the ensemble classifier are better than the \texttt{k}-NN methods, but not as high as the NeuN classifier.   Among the training sets, only Set C identifies all 5 known dwarfs in all the methods. The other two sets work well in all classifiers except \texttt{k}-NN-TD. The classification methods using 3-class training sets also work extremely well (Table \ref{table:Regions3class}), identifying all 5 known dwarfs with a \textit{RR} around 99.8\%.

\subsection{Hercules}
The selected region in Hercules contains 2,611 objects with photometry in all the requisite bands. The region is known to contain 3 known brown dwarfs. Application of the classification methods towards the objects in this region show that among the 2-class methods, NeuN classifier (Table~\ref{table:Regions2class}) again produces the best results with two out of three training sets identifying all known brown dwarfs, all with \textit{RR} values above 99.5\%. This is followed by \texttt{k}-NN-C and \texttt{k}-NN-TD methods. In this region, the ensemble classifier performs better than the \texttt{k}-NN methods, but not as well as the NeuN classifier. Among the training sets, only Set C identifies all 3 known dwarfs and has a \textit{RR} value above 99\%. Training set A performs well in the NeuN (identifying all 3 known dwarfs with \textit{RR}=99.53\%) but is unable to identify the known dwarfs by the other two methods. The classification techniques using 3-class training sets perform well (Table~\ref{table:Regions3class}), identifying all 3 known dwarfs with a \textit{RR} value $\sim99.8$\%.

\subsection{Lyra}
The selected region in the constellation Lyra has 4,620 objects with photometry in all infrared wavebands considered here. No known dwarf is found in this region. In this region, the NeuN and the ensemble classifiers (Table \ref{table:Regions2class}) work well giving \textit{RR} above 99\% for all training sets. The \texttt{k}-NN-C method does not fare so well in comparison, with only sets B and C having a \textit{RR} of above 99\%. The same follows for \texttt{k}-NN-TD method, with sets A and B having a \textit{RR} of above 99\%. But both the \texttt{k}-NN methods have \textit{RR}>96\% for all sets. The 3-class classification techniques (Table \ref{table:Regions3class}) give a high \textit{RR}, but classify an abnormally large number of objects as L-dwarfs, which may be due to the inclusion of early L-dwarfs from \citet{Best2018} in the 3-class training set.

\begin{table*}
	\centering
	\caption{Results on the three regions in sky for 2-class training sets.}
	\begin{tabular}{ccccrccrccr}
		\hline\hline
		\multicolumn{2}{c}{Training Sets} & \multicolumn{3}{c}{Set A} & \multicolumn{3}{c}{Set B} & \multicolumn{3}{c}{Set C} \\ \hline
		Region & Method & N$^1$ & KR$^2$ & \begin{tabular}[r]{@{}r@{}}RR$^3$\\ (\%)\end{tabular} & N$^1$ & KR$^2$ & \begin{tabular}[c]{@{}c@{}}RR$^3$\\ (\%)\end{tabular} & N$^1$ & KR$^2$ & \begin{tabular}[r]{@{}r@{}}RR$^3$\\ (\%)\end{tabular} \\ \hline
		& NeuN & 16 & 5/5 & 99.9 & 7 & 5/5 & 99.9 & 20 & 5/5 & 99.9 \\
		& \texttt{k}-NN-C (\texttt{k}=5) & 351 & 5/5 & 98.3 & 68 & 5/5 & 99.7 & 510 & 5/5 & 97.6 \\
		Serpens & \texttt{k}-NN-C (\texttt{k}=10) & 562 & 4/5 & 97.3 & 78 & 5/5 & 99.6 & 123 & 5/5 & 99.4 \\
		& \texttt{k}-NN-TD (\texttt{k}=5) & 45 & 0/5 & 99.8 & 1 & 0/5 & 100 & 1230 & 5/5 & 94.2 \\
		& \texttt{k}-NN-TD (\texttt{k}=10) & 25 & 0/5 & 99.9 & 1 & 0/5 & 100 & 1294 & 5/5 & 93.8 \\
		& Ensemble & 47 & 4/5 & 99.8 & 8 & 5/5 & 99.9 & 27 & 5/5 & 99.9 \\ \hline
		& NeuN & 4 & 3/3 & 99.9 & 3 & 2/3 & 99.9 & 5 & 3/3 & 99.8 \\
		& \texttt{k}-NN-C (\texttt{k}=5) & 22 & 2/3 & 99.2 & 5 & 2/3 & 99.8 & 72 & 3/3 & 97.2 \\
		Hercules & \texttt{k}-NN-C (\texttt{k}=10) & 37 & 2/3 & 98.6 & 7 & 2/3 & 99.7 & 19 & 3/3 & 99.3 \\
		& \texttt{k}-NN-TD (\texttt{k}=5) & 4 & 1/3 & 99.9 & 1 & 1/3 & 99.9 & 69 & 3/3 & 97.4 \\
		& \texttt{k}-NN-TD (\texttt{k}=10) & 2 & 1/3 & 99.9 & 1 & 1/3 & 99.9 & 35 & 3/3 & 98.7 \\
		& Ensemble & 3 & 1/3 & 99.9 & 1 & 1/3 & 99.9 & 7 & 3/3 & 99.7 \\ \hline
		& NeuN & 7 & - & 99.8 & 1 & - & 99.9 & 2 & - & 99.9 \\
		& \texttt{k}-NN-C (\texttt{k}=5) & 81 & - & 98.3 & 22 & - & 99.5 & 139 & - & 97.0 \\
		Lyra & \texttt{k}-NN-C (\texttt{k}=10) & 119 & - & 97.4 & 23 & - & 99.5 & 46 & - & 99.0 \\
		& \texttt{k}-NN-TD (\texttt{k}=5) & 9 & - & 99.8 & 0 & - & 100 & 149 & - & 96.8 \\
		& \texttt{k}-NN-TD (\texttt{k}=10) & 6 & - & 99.9 & 2 & - & 99.9 & 162 & - & 96.5 \\
		& Ensemble & 9 & - & 99.8 & 1 & - & 99.9 & 9 & - & 99.8 \\ \hline \hline
	\end{tabular}

         $^1$N - Number of brown dwarfs \\
         $^2$KR - Ratio of known brown dwarfs \\
         $^3$RR - Rejection Ratio
	\label{table:Regions2class}
\end{table*}



\begin{table*}
	\centering
	\caption{Results on the three regions in sky for 3-class training sets.}
	\begin{tabular}{ccccccr}
		\hline\hline
		Region & Method & \begin{tabular}[c]{@{}c@{}}No of objects\\ in class Bg\end{tabular} & \begin{tabular}[c]{@{}c@{}}No of objects\\ in class L\end{tabular} & \begin{tabular}[c]{@{}c@{}}No of objects\\ in class T\&Y\end{tabular} & KR$^2$ 
            &  \begin{tabular}[r]{@{}r@{}}RR$^3$\\ (\%)\end{tabular}  \\ \hline
		& NeuN & 18354 & 2629 & 39 & 5/5 & 99.8 \\
		Serpens & \texttt{k}-NN-C (\texttt{k}=5) & 15827 & 5172 & 23 & 5/5 & 99.9 \\
		& \texttt{k}-NN-C (\texttt{k}=10) & 2218 & 389 & 4 & 5/5 & 99.9 \\ \hline
		& NeuN & 2395 & 208 & 8 & 3/3 & 99.7 \\
		Hercules & \texttt{k}-NN-C (\texttt{k}=5) & 2196 & 411 & 4 & 3/3 & 99.9 \\
		& \texttt{k}-NN-C (\texttt{k}=10) & 2218 & 389 & 4 & 3/3 & 99.9 \\ \hline
		& NeuN & 3713 & 541 & 6 & - & 99.9 \\
		Lyra & \texttt{k}-NN-C (\texttt{k}=5) & 3183 & 1071 & 6 & - & 97.0 \\
		& \texttt{k}-NN-C (\texttt{k}=10) & 3300 & 955 & 5 & - & 99.0 \\ \hline\hline
	\end{tabular}

         $^2$KR - Ratio of known brown dwarfs \\
         $^3$RR - Rejection Ratio
	\label{table:Regions3class}
\end{table*}


\subsection{Search for counterparts}

\subsubsection{SIMBAD counterparts}

The objects identified as brown dwarf candidates by the NeuN and ensemble classifiers for each region are given in Tables \ref{table:Simbad1}, \ref{table:Simbad2}, \ref{table:Simbad3}. We carried out a search in the SIMBAD astronomical database to see if these sources matched with any known objects identified previously. The search was applied based on the positional coordinates of the source and a search radius of $3.5''$. This corresponds to half the beam of the W3 WISE band, which has the largest beams among the bands considered.   

In the Serpens region, both the NeuN and ensemble classifiers together identify 59 objects, out of which 9 objects have counterparts in the SIMBAD database. We were pleasantly surprised to see one of them as a L-type brown dwarf , 2MASS J16192830+0050118. This was not included in the previous 5 known dwarfs based on the catalogs considered, as its corresponding WISE data was not linked to this particular object.  The other 8 objects are M-type variable stars and galaxies. In the Hercules region, out of the 6 candidates, only one has a previously identified counterpart (an M-type variable star). The classifiers identify 15 brown dwarf candidates in the Lyra region, of which four have SIMBAD counterparts. Two of these are variable stars and two are radio sources. While the association provides a certain estimate of the type of source, one caveat is that due to the WISE resolution, the search diameter considered for the positional search is not small, i.e. $7''$. 
		
\subsubsection{Gaia Counterparts}

A search was also carried out using the Gaia database using the DR2 catalog \citep{2018A&A...616A...1G} for the objects identified as brown dwarf candidates by the NeuN and ensemble classifiers. Again, a search radius of $3.5''$ was used and the nearest positional association was considered. For every source which had a corresponding Gaia identification, the parallaxes, proper motions and photometric magnitudes were obtained. The distance to each source was derived from the parallax. Using the distance and apparent G magnitude, we estimated absolute G magnitude of each source. This was then was compared with the absolute magnitudes of known L and T brown dwarfs \citep{Best2018} to estimate the spectral type of each source. The associated Gaia sources and their properties are listed in Table~\ref{table:Gaia}.

Three objects in Hercules, sixteen in Serpens and two in Lyra were found to have Gaia associations. Of these, only one object is of spectral type L or later and that is the known L-type brown dwarf  2MASS J16192830+0050118 which was identified earlier from the SIMBAD database. It is also the only object which is within a distance of 100 parsecs from the Earth. The majority of sources do not have Gaia counterparts. This is expected as the cooler brown dwarfs are likely to be faint at optical wavelengths.    


\begin{table*}
       \scriptsize
	\centering
	\caption{Brown dwarf candidates identified by NeuN and Ensemble Classifier in Hercules, the alphabets in brackets indicate the training set.}
\begin{tabular}{crrccc}
			\hline\hline
			WISE ID & \begin{tabular}[l]{@{}l@{}}$\alpha_{\rm{J2000}}$\\ (deg)\end{tabular}  & \begin{tabular}[c]{@{}c@{}}$\delta_{\rm{J2000}}$\\ (deg)\end{tabular}  & Technique & \begin{tabular}[c]{@{}c@{}}SIMBAD\\ Association\end{tabular} & Category \\ \hline
			J175510.28+180320.2 & 268.7928 & 18.0555 & NeuN(A,B,C); Ensemble(A,B,C) & 2MASS J17551062+1803203 & Brown Dwarf \\
			J175032.93+175904.2 & 267.6378 & 17.9844 & NeuN(A,C); Ensemble(C) & 2MASS J17503293+1759042 & Brown Dwarf \\
			J175454.47+164919.6 & 268.7273 & 16.8218 & NeuN(A,B,C); Ensemble(C) & 2MASS J17545447+1649196 & Brown Dwarf \\
			J174846.58+165936.8 & 267.1941 & 16.9935 & NeuN(C); Ensemble(A,C) & No object found &  \\
			J175527.28+162322.9 & 268.8637 & 16.3897 & Ensemble(A) & No object found &  \\
			J175338.61+164013.8 & 268.4109 & 16.6705 & NeuN(A); Ensemble(C) & NSV 9827 & Semi-regular pulsating star \\
			J174811.68+174109.6 & 267.0487 & 17.6860 & NeuN(C) & No object found &  \\
			J174827.01+175113.4 & 267.1125 & 17.8537 & Ensemble(C) & No object found &  \\
			J174854.84+164232.2 & 267.2285 & 16.7089 & Ensemble(C) & No object found & \\ \hline\hline
		\end{tabular}
		\label{table:Simbad1}
	\end{table*}


	\begin{table*}
               \scriptsize
		\centering
		\caption{Brown dwarf candidates identified by NeuN and Ensemble Classifier in Lyra, the alphabets in brackets indicate the training set.}
		\begin{tabular}{crrccc}
				\hline\hline
				WISE ID & \begin{tabular}[c]{@{}c@{}}$\alpha_{\rm{J2000}}$\\ (deg)\end{tabular}  & \begin{tabular}[c]{@{}c@{}}$\delta_{\rm{J2000}}$\\ (deg)\end{tabular}  & Technique & \begin{tabular}[c]{@{}c@{}}SIMBAD\\ Association\end{tabular} & Category \\ \hline
				J181811.58+325942.6 & 274.5483 & 31.5300 & Ensemble(C) & No object found &  \\
				J181820.10+314200.5 & 274.5837 & 31.7001 & NeuN(A,B,C); Ensemble(A,B,C) & V* A0 Lyr	& Variable star of Mira Cet type\\
				J181850.55+315116.8 & 274.7106 & 31.8546 & NeuN(A); Ensemble(A) & No object found &  \\
				J181811.58+325942.6 & 274.8476 & 31.7453 & Ensemble(C) & No object found &  \\
				J181950.91+325747.9 & 274.9621 & 32.9633 & NeuN(A); Ensemble(A,C) & No object found &  \\
				J182043.02+314006.9 & 275.1793 & 31.6686 & Ensemble(A) & NVSS J182043+314006 &	Radio Source  \\
				J182046.97+332431.9	& 275.1957 & 33.4089 & Ensemble(C) & No object found & \\
				J182047.58+313841.0 & 275.1983 & 31.6447 & Ensemble(C) & No object found &  \\
				J182138.94+323907.4 & 275.4122 & 32.6521 & NeuN(A) & No object found &  \\
				J182240.79+332256.1 & 275.6699 & 33.3822 & Ensemble(A) & No object found &  \\
				J182246.23+322258.9 & 275.6926 & 32.3830 & NeuN(A); Ensemble(A,C) & NVSS J182246+322258	& Radio Source  \\
			    J182259.75+314404.1	& 275.7490 & 31.7345 & Ensemble(A) & No object found &  \\
				J182302.37+321010.3 & 275.7599 & 32.1695 & Ensemble(C) & No object found &  \\
				J182528.82+313045.4 & 276.3701 & 31.5126 & NeuN(A); Ensemble(A) & No object found &  \\
				J182529.61+313305.3 & 276.3734 & 31.5515 & NeuN(A,C); Ensemble(A,C) & V* IS Lyr	& Variable star of Mira Cet type	  \\ \hline\hline
			\end{tabular}
		\label{table:Simbad2}
		\end{table*}



		\begin{table*}
			\scriptsize
			\centering
			\caption{Brown dwarf candidates identified by NeuN and Ensemble Classifier in Serpens, the alphabets in brackets indicate the training set.}
			{\begin{tabular}{crrccc}
					\hline\hline
					WISE ID & \begin{tabular}[c]{@{}c@{}}$\alpha_{\rm{J2000}}$\\ (deg)\end{tabular}  & \begin{tabular}[c]{@{}c@{}}$\delta_{\rm{J2000}}$\\ (deg)\end{tabular}  & Technique & \begin{tabular}[c]{@{}c@{}}SIMBAD\\Association\end{tabular} & Category \\ \hline
					J162918.56+033535.5 & 247.3273 & 3.5932 &  NeuN(A,B,C); Ensemble(A,B,C) & 2MASS J16291840+0335371 & Brown Dwarf  \\
					J163236.47+032927.3 & 248.1520 & 3.4909 &  NeuN(A,B,C); Ensemble(A,B,C) & 2MASS J16323642+0329269 & Brown Dwarf  \\
					J161927.53+031348.1 & 244.8649 & 3.2299 &  NeuN(A,B,C); Ensemble(A,B,C) & 2MASS J16192751+0313507 &  Brown Dwarf binary \\
					J162414.07+002915.6 & 246.0587 & 0.4877 &  NeuN(A,B,C); Ensemble(A,B,C) & 2MASS J16241436+0029158  & Brown Dwarf \\
					J163022.90+081821.0 & 247.5954 & 8.3058 &  NeuN(A,B,C); Ensemble(B,C) & 2MASS J16302295+0818221 & Brown Dwarf \\
					J161800.81+074046.4 & 244.5034 & 7.6795 & NeuN(A) & No object found &  \\
					J161804.33+034225.6 & 244.5180 & 3.7071 & Ensemble(C) & No object found &  \\
					J161817.46+073151.3 & 244.5728 & 7.5309 & Ensemble(A) & No object found &  \\
					J161835.18+070629.7 & 244.6466 & 7.1083 & Ensemble(A) & No object found &  \\
					J161910.21+011905.8 & 244.7926 & 1.3183 & NeuN(C); Ensemble(C) & No object found &  \\
					J161910.47+064223.3 & 244.7936 & 6.7065 & NeuN(C); Ensemble(C) & SDSS J161910.48+064223.2 & Starburst Galaxy \\
					J161915.97+083555.5 & 244.8166 & 8.5987 & Ensemble(A) & No object found &  \\
					J161917.17+060856.5 & 244.8215 & 6.1490 & Ensemble(A) & No object found &  \\
					J161928.33+005010.8 & 244.8681 & 0.8363 & NeuN(A) & 2MASS J16192830+0050118 & Brown Dwarf \\
					J161938.90+043203.8 & 244.9121 & 4.5344 & NeuN(C); Ensemble(C) & No object found &  \\
					J161950.93+082348.4 & 244.9622 & 8.3968 & Ensemble(A) & No object found &  \\
					J162004.13+002141.9 & 245.0172 & 0.3616 & Ensemble(A) & No object found &  \\
					J162017.70+040808.8 & 245.0737 & 4.1358 & Ensemble(A) & No object found &  \\
					J162037.66+014617.6 & 245.1569 & 1.7715 & NeuN(A) & No object found &  \\
					J162055.66+070510.0 & 245.2319 & 7.0861 & Ensemble(A) & No object found &  \\
					J162106.03+073555.9 & 245.2751 & 7.5988 & Ensemble(C) & No object found &  \\
					J162208.71+014252.2 & 245.5363 & 1.7145 & NeuN(C); Ensemble(C) & No object found &  \\
					J162221.38+030814.8 & 245.5891 & 3.1374 & Ensemble(A) & No object found &  \\
					J162308.40+002259.7 & 245.7850 & 0.3832 & Ensemble(A) & No object found &  \\
					J162327.26+000643.4 & 245.8636 & 0.1120 & Ensemble(C) & No object found &  \\
					J162348.61+043008.2 & 245.9526 & 4.5023 & NeuN(C);  Ensemble(C) & No object found &  \\
					J162350.29+053635.4 & 245.9595 & 5.6098 & NeuN(A) & No object found &  \\
					J162359.75+082017.6 & 245.9990 & 8.3382 & NeuN(A) & No object found &  \\
					J162407.87+072619.7 & 246.0328 & 7.4388 & Ensemble(C) & No object found &  \\
					J162426.65+054218.5 & 246.1110 & 5.7051 & Ensemble(A) & No object found &  \\
					J162450.40+054231.4 & 246.2100 & 5.7087 & Ensemble(A) & No object found &  \\
					J162450.47+014948.1 & 246.2103 & 1.8300 & Ensemble(A) & No object found &  \\
					J162454.89+025957.7 & 246.2287 & 2.9994 & NeuN(C) & V* V911 Oph & Long-period variable star \\
					J162532.03+012123.7 & 246.3834 & 1.3566 & Ensemble(A) & No object found &  \\
					J162532.15+055407.0 & 246.3839 & 5.9019 & Ensemble(C) & No object found &  \\
					J162616.28+051700.6 & 246.5678 & 5.2835 & Ensemble(A) & No object found &  \\
					J162618.85+041127.6 & 246.5785 & 4.1910 & NeuN(A) & No object found &  \\
					J162623.15+020009.7 & 246.5965 & 2.0027 & Ensemble(C) & No object found &  \\
					J162642.50+045429.5 & 246.6771 & 4.9082 & Ensemble(A) & No object found &  \\
					J162655.81+083510.7 & 246.7325 & 8.5863 & Ensemble(A) & No object found &  \\
					J162716.25+004031.8 & 246.8177 & 0.6755 & Ensemble(A) & No object found &  \\
					J162810.89+003947.7 & 247.0454 & 0.6632 & Ensemble(C) & No object found &  \\
					J162815.15+005836.8 & 247.0631 & 0.9769 & NeuN(B,C); Ensemble(C) & No object found &  \\
					J162823.78+072620.6 & 247.0991 & 7.4390 & NeuN(C); Ensemble(C) & No object found &  \\
					J162903.50+055218.3 & 247.2646 & 5.8717 & NeuN(A,C); Ensemble(C) & No object found &  \\
					J162936.12+060416.8 & 247.4005 & 6.0713 & NeuN(C); Ensemble(C) & No object found &  \\
					J162948.11+054951.4 & 247.4504 & 5.8309 & NeuN(A) & No object found &  \\
					J163030.11+030044.7 & 247.6255 & 3.0124 & Ensemble(A) & No object found &  \\
					J163045.42+065252.6 & 247.6893 & 6.8813 & NeuN(C); Ensemble(C) & SDSS J163045.43+065252.5 & Blue object \\
					J163053.87+045609.3 & 247.7244 & 4.9359 & Ensemble(C) & LARCS a2204r04-2748 & Galaxy in a cluster \\
					J163109.27+033600.5 & 247.7887 & 3.6001 & NeuN(C); Ensemble(B,C) & No object found &  \\
					J163114.37+044720.4 & 247.8099 & 4.7890 & NeuN(C); Ensemble(C) & No object found &  \\
					J163157.59+030610.9 & 247.9899 & 3.1030 & NeuN(C); Ensemble(B,C) & V* V721 Oph & Variable star (Mira Cet type) \\
					J163236.72+014649.0 & 248.1530 & 1.7803 & Ensemble(A) & No object found &  \\
					J163240.24+055427.9 & 248.1677 & 5.9077 & Ensemble(A) & No object found &  \\
					J163247.31+021041.2 & 248.1971 & 2.1781 & Ensemble(A) & No object found &  \\
					J163248.85+073420.3 & 248.2035 & 7.5723 & Ensemble(A) & No object found &  \\
					J163255.54+065129.7 & 248.2314 & 6.8582 & NeuN(A,B,C); Ensemble(B,C) & V* SS Her & Variable star (Mira Cet type) \\
					J163304.98+051652.0 & 248.2707 & 5.2811 & Ensemble(A) & No object found &  \\
					J163314.56+011850.4 & 248.3106 & 1.3140 & Ensemble(A) & No object found &  \\
					J163343.00+064010.3 & 248.4291 & 6.6695 & Ensemble(A) & No object found &  \\
					J163345.48+070903.2 & 248.4395 & 7.1509 & Ensemble(A) & PMN J1633+0708 & Radio source \\
					J163351.17+002959.7 & 248.4632 & 0.4999 & Ensemble(A) & 2MASX J16335114+0029595 & Galaxy \\
					J163356.21+062654.2 & 248.4842 & 6.4484 & Ensemble(A) & No object found &  \\ \hline\hline
				\end{tabular}}
			\label{table:Simbad3}	
			\end{table*}


\begin{table*}
       \scriptsize
	\centering
	\caption{The Gaia associations and their properties (parallaxes, proper motions and estimates spectral types), of the brown dwarf candidates identified by NeuN and Ensemble Classifier in all three regions.}
	\begin{tabular}{crrccccc}
		\hline\hline
		Region   & \begin{tabular}[c]{@{}c@{}}$\alpha_{\rm{J2000}}$\\ (deg)\end{tabular}        & \begin{tabular}[c]{@{}c@{}}$\delta_{\rm{J2000}}$\\ (deg)\end{tabular}     & Gaia Source Id      & \begin{tabular}[c]{@{}c@{}}Parallax\\ (mas)\end{tabular} & \begin{tabular}[c]{@{}c@{}}Distance\\ (parsecs)\end{tabular} & \begin{tabular}[c]{@{}c@{}}Estimated\\ Spectral\\ Type$^1$\end{tabular} & \begin{tabular}[c]{@{}c@{}}Proper Motion\\ (mas/yr)\end{tabular} \\ \hline
		Hercules & 267.1941 & 16.9935 & 4550887629617161344 & 0.0972                                                      & 10284.8809                                                   & \textgreater L0                                                     & 0.172097                                                         \\
		& 267.0487 & 17.686  & 4550974216159379200 & 0.4213                                                      & 2373.1620                                                    & \textgreater L0                                                     & 4.523704                                                         \\
		& 267.1125 & 17.8537 & 4551012973945004032 & 0.6534                                                      & 1530.2454                                                    & \textgreater L0                                                     & 5.065651                                                         \\ \hline
		Serpens  & 244.5181 & 3.7071  & 4436379751351585536 & 6.0044                                                      & 166.5441                                                     & \textgreater L0                                                     & 21.157053                                                        \\
		& 244.7926 & 1.3183  & 4408885737249657216 & 0.9791                                                      & 1021.2935                                                    & \textgreater L0                                                     & 6.5097901                                                        \\
		& 244.8681 & 0.8363  & 4408577118080317056 & 37.2891                                                     & 26.8175                                                      & L0 - L2                                                             & 96.588497                                                        \\
		& 245.0737 & 4.1358  & 4436510696314413312 & 1.6612                                                      & 601.9658                                                     & \textgreater L0                                                     & 0.726790                                                         \\
		& 245.5363 & 1.7145  & 4432757204136317440 & 0.1174                                                      & 8515.7912                                                    & \textgreater L0                                                     & 1.553490                                                         \\
		& 246.2287 & 2.9994  & 4433148282382642304 & 0.6089                                                      & 1642.3517                                                    & \textgreater L0                                                     & 6.897879                                                         \\
		& 246.3839 & 5.9019  & 4438657282313809408 & 0.4231                                                      & 2363.6879                                                    & \textgreater L0                                                     & 8.115868                                                         \\
		& 247.0991 & 7.4390  & 4439633854803047680 & 0.4901                                                      & 2040.3025                                                    & \textgreater L0                                                     & 0.436674                                                         \\
		& 247.4005 & 6.0713  & 4438724112005249152 & 1.2443                                                      & 803.6814                                                     & \textgreater L0                                                     & 0.944441                                                         \\
		& 247.6893 & 6.8813  & 4438837636580679680 & 0.0376                                                      & 26602.8631                                                   & \textgreater L0                                                     & 0.340288                                                         \\
		& 247.7886 & 3.6001  & 4433462811427426816 & 0.3383                                                      & 2955.5194                                                    & \textgreater L0                                                     & 8.287301                                                         \\
		& 248.1677 & 5.9077  & 4438472461282781696 & 4.1644                                                      & 240.1279                                                     & \textgreater L0                                                     & 31.395639                                                        \\
		& 248.1971 & 2.1781  & 4432414564529967232 & 2.3157                                                      & 431.8293                                                     & \textgreater L0                                                     & 2.881503                                                         \\
		& 248.2314 & 6.8582  & 4439156220075003520 & 0.5491                                                      & 1821.0399                                                    & \textgreater L0                                                     & 6.037068                                                         \\
		& 248.2707 & 5.2811  & 4435337375673394944 & 0.1840                                                      & 5433.4089                                                    & \textgreater L0                                                     & 0.390203                                                         \\
		& 248.4395 & 7.1509  & 4439197211243356672 & 0.4347                                                      & 2300.4445                                                    & \textgreater L0                                                     & 0.807281                                                         \\ \hline
		Lyra     & 274.5483 & 31.5300 & 4591800083627498496 & 0.3162                                                      & 3162.5501                                                    & \textgreater L0                                                     & 9.983660                                                         \\
		& 275.4122 & 32.6521 & 4592724906048223616 & 0.3433                                                      & 2912.5313                                                    & \textgreater L0                                                     & 6.467676                                                         \\ \hline\hline
	\end{tabular}
	\label{table:Gaia}

	$^1$~$>$ means earlier spectral type.
\end{table*}


\section{Inferences}
\label{sec:Sec6}
The NeuN classifier emerges as the best technique out of the individual methods, performing well in training set efficiency calculations as well as on tests in regions on the sky. The \texttt{k}-NN-C method with 3-class training set also holds promise with $\mathcal{R}$  for T-dwarfs as high as 99\%, and $\mathcal{C}$ values in the range 90-95\%. NeuN gives much better results than \texttt{k}-NN-C with 2-class training set while testing on specific regions in the sky. This method also identified a brown dwarf in the Serpens region which was not part of the initial training sets or the dwarfs identified by WISE. The \texttt{k}-NN-C using 2-class training set gives results comparable to NeuN while comparing training set efficiencies but does not hold up as much while testing on known test samples or new regions. The \texttt{k}-NN-TD method fares less well than the previously discussed methods in the training set efficiency calculations and the Hercules test region.

Thus, if the classification is to be implemented using a single method, the NeuN classifier
would be an appropriate choice. But a better option would be to use the ensemble classifier, which performs reasonably well in all scenarios, and where the decision would not be dependent on a single classifier alone. Of all the training sets used (in 2-class classifications), training set C performs best on sources from given regions on the sky. It shows maximum efficiency in the Hercules and Serpens regions, and high rejection ratios in the Lyra region. But the set has low training set efficiency in cross-validation (one amongst the lowest in both \texttt{k}-NN methods). Training Set B was found to be effective in rejecting background sources but its performance in the Hercules region was poor compared to training set C. Training set A has high efficiencies and managed to identify the new brown dwarf in the Serpens region, but it does not hold up as well in the other aspects. Also, it has the least sample variety among the 3 training sets. The fact that Set C, which has the most sample variety, is also the best performer seems to indicate that there is a strong correlation between performance and background sample variety. Set B has intermediate sample variety and its performance is also intermediate that of the other 2 sets. Since astronomical objects are found in a vast variety, inclusion of different types of background objects helps machine-learning algorithms perform better.  

A number of objects have been identified in the three regions using the classification techniques described in the work. We note that the SIMBAD associations indicate that we are also selecting objects which are not brown dwarfs, viz. M-type stars, few galaxies and carbon stars. In one case, a radio source is also identified as a counterpart. Thus, one needs to probe in detail in order to confirm the associations of the identified candidates. However, the fact that known brown dwarfs have been identified by these methods provides strong support to the fact that the methods are effective and one can verify the nature of the other brown dwarf `candidates' through follow-up spectroscopic studies. It is worth noting that the number of brown-dwarf candidates identified by these classifiers is much lower than those identified by traditional techniques of colour-magnitude restrictions which, in turn, saves time and resources required for the final verification.

We have used WISE and 2MASS data for this study as they were easily available for numerous different objects. A more robust classification calls for additional information in the form of  identification of the background objects, eg. YSOs in a given star-forming region. Alternate examples include cross-identifications of background stars and galaxies across catalogs, or investigation of regions away from Galactic plane where the number density of stars is lower than the Galactic plane. The amount and nature of extinction towards each object is also expected to play a crucial role in ascribing a background source as a brown dwarf.
This additional information provided to the classifiers can improve the classification and reliability of the methods. Lastly, we note that these classification techniques can be used to identify any group of elusive astronomical objects, by changing the input colours and training sets. Hence, this approach can serve as a base for classification of other astronomical objects as well.

\section{Summary}
\label{sec:Sec7}
\begin{itemize}
	\item NeuN and \texttt{k}-NN methods  have been used for classifying astronomical objects based on their photometric colours. Although the methods are general and can be applied to select any specific kind of astronomical objects, we have applied it to the specific case of brown dwarfs.
	\item In this study, apart from NeuN, we have used two different \texttt{k}-NN methods: \texttt{k}-NN-C and \texttt{k}-NN-TD, for classifying brown dwarfs, using six colours from WISE and 2MASS. We also propose an ensemble classifier which  identifies brown dwarf candidates on the basis of a majority vote from the above three methods.
	\item A number of training sets have been constructed for testing the performance of the classifiers. This includes the 2-class and 3-class training sets.
	\item In addition to the different techniques, we create different training sets by combining templates from various known brown dwarf and background object catalogs. The efficiencies for the sets, and for different methods, are then calculated by using $\mathcal{C}$ and $\mathcal{R}$ as the validation metrics.
	\item All the methods perform well on the training sets considered, with $\mathcal{C} \ge 87\%$ and $\mathcal{R} \ge 88\%$. In the 2-class classification, both NeuN and the ensemble classifier emerge as the best methods. Both NeuN and \texttt{k}-NN-C perform equally well in the 3-class clssification methods.
	\item We apply the methods and optimal training sets to three regions in the sky: Serpens, Hercules and Lyra. Of these, Serpens and Hercules have known brown dwarfs, previously identified by WISE. 
	\item The NeuN classifier performs relatively better than the \texttt{k}-NN methods in the three regions, in the 2-class classification, identifying all the previously known dwarfs. This is followed by the ensemble classifier. The two \texttt{k}-NN methods do not fare as well, with\texttt{k}-NN-C being the better of the two. 
	\item The 3-class classification also holds promise with its performance equalling or even exceeding that of the 2-class NeuN.
	\item A search for counterparts in the SIMBAD and Gaia databases was also carried out for the brown dwarf candidates from each region. This led to the identification of one of the candidates in the Serpens region as a brown dwarf which was not part of the brown dwarfs identified by WISE. A fraction of the other candidates are variable stars and other background objects. 
	\item These methods of multi-dimensional classification based on photometric colours are expected to significantly downsize the candidate sample for follow-up studies, as compared to traditional colour and magnitude diagrams or threshold cuts. 
\end{itemize}

\section*{Acknowledgements}

We thank the referee M. Marengo for useful suggestions that have improved the paper presentation. This publication makes use of data products from the Wide-Field Infrared Survey Explorer,
which is a joint project of the University of California, Los Angeles, and the Jet Propulsion
Laboratory/California Institute of Technology, funded by the National Aeronautics and Space
Administration. This publication also makes use of data products from 2MASS, which is a joint project of the University of Massachusetts and the Infrared Processing and Analysis Center/California Institute of Technology, funded by the National Aeronautics and Space Administration and the National Science Foundation. This research has made use of the Vizier and SIMBAD databases, operated at CDS, Strasbourg, France.This work has also benefitted from the M, L, and T dwarf compendium housed at DwarfArchives.org, whose server was funded by a NASA Small Research Grant, administered by the American Astronomical Society. This work presents results from the European Space Agency (ESA) space mission Gaia (http://www.cosmos.esa.int/gaia) taken from the archive website https://archives.esac.esa.int/gaia. The data was processed by the Gaia Data Processing and Analysis Consortium (DPAC) which is funded by national institutions, in particular the institutions participating in the Gaia MultiLateral Agreement (MLA).





\bibliographystyle{mnras}
\bibliography{BDclass_ref} 








\bsp	
\label{lastpage}
\end{document}